\documentclass[12pt]{article}
\usepackage{epsfig,cite}
\usepackage {amsmath}
\usepackage{amssymb}
\include{epsf}
\newcount\timecount
\newcount\hours \newcount\minutes  \newcount\temp \newcount\pmhours
\hours = \time
\divide\hours by 60
\temp = \hours
\multiply\temp by 60
\minutes = \time
\advance\minutes by -\temp
\def\hour{\the\hours}
\def\minute{\ifnum\minutes<10 0\the\minutes
            \else\the\minutes\fi}
\def\clock{
\ifnum\hours=0 12:\minute\ AM
\else\ifnum\hours<12 \hour:\minute\ AM
      \else\ifnum\hours=12 12:\minute\ PM
            \else\ifnum\hours>12
                 \pmhours=\hours
                 \advance\pmhours by -12
                 \the\pmhours:\minute\ PM
                 \fi
            \fi
      \fi
\fi
}

\def\monthname{\relax\ifcase\month 0/\or January\or February\or
   March\or April\or May\or June\or July\or August\or September\or
   October\or November\or December\else\number\month/\fi}

\def\bold#1{\setbox0=\hbox{$#1$}%
     \kern-.025em\copy0\kern-\wd0
     \kern.05em\copy0\kern-\wd0
     \kern-.025em\raise.0433em\box0 }

\def\beq{\begin{equation}}
\def\eeq{\end{equation}}

\def\ga{\mathrel{\raise.3ex\hbox{$>$\kern-.75em\lower1ex\hbox{$\sim$}}}}
\def\la{\mathrel{\raise.3ex\hbox{$<$\kern-.75em\lower1ex\hbox{$\sim$}}}}
\def\gev{{\rm \, Ge\kern-0.125em V}}
\def\tev{{\rm \, Te\kern-0.125em V}}
\def\gyr{{\rm \, G\kern-0.125em yr}}

\def\tbt{\tan \beta}
\def\ttbt{\tan^2 \beta}

\def\gappeq{\mathrel{\rlap {\raise.5ex\hbox{$>$}}
{\lower.5ex\hbox{$\sim$}}}}
\def\lappeq{\mathrel{\rlap{\raise.5ex\hbox{$<$}}
{\lower.5ex\hbox{$\sim$}}}}
\def\Toprel#1\over#2{\mathrel{\mathop{#2}\limits^{#1}}}

\def\stau{\widetilde \tau}
\def\stop{\widetilde t}

\def\m12{m_{1\!/2}}

\def\stau{\tilde{\tau}}

\def\bea{\begin{eqnarray}}
\def\eea{\end{eqnarray}}
\newcommand{\goto}{\rightarrow}
\newcommand{\bmm}{B_s \goto \mu^+ \, \mu^-}

\begin{document}
\begin{titlepage}
\pagestyle{empty}
\baselineskip=21pt
\rightline{\tt hep-ph/yymmnnn}
\rightline{CERN-PH-TH/2007-073}
\rightline{UMN--TH--2602/07}
\rightline{FTPI--MINN--07/15}
\vskip 0.2in
\begin{center}
{\large{\bf Phenomenology of GUT-less Supersymmetry Breaking}}
\end{center}
\begin{center}
\vskip 0.2in
{\bf John~Ellis}$^1$, {\bf Keith~A.~Olive}$^{2}$ and
{\bf Pearl Sandick}$^{2}$
\vskip 0.1in

{\it
$^1${TH Division, PH Department, CERN, CH-1211 Geneva 23, Switzerland}\\
$^2${William I. Fine Theoretical Physics Institute, \\
University of Minnesota, Minneapolis, MN 55455, USA}\\
}

\vskip 0.2in
{\bf Abstract}
\end{center}
\baselineskip=18pt \noindent

We study models in which supersymmetry breaking appears at an intermediate scale, $M_{in}$, below the GUT scale.  We assume that the soft supersymmetry-breaking parameters of the MSSM are 
universal at $M_{in}$, and analyze the morphology of the constraints from cosmology and collider experiments on the allowed regions of parameter space as $M_{in}$ is reduced from the GUT scale.  We present separate analyses of the $(m_{1/2},m_0)$ planes for $\tbt=10$ and $\tbt=50$, as well as a discussion of non-zero trilinear couplings, $A_0$.  Specific scenarios where the gaugino and scalar masses appear to be universal below the GUT scale have been found in mirage-mediation models, which we also address here. We demand that the lightest neutralino be the LSP, and that the relic neutralino density
not conflict with measurements by WMAP and other observations. 
At moderate values of $M_{in}$, we find that the
allowed regions of the $(m_{1/2},m_0)$ plane are squeezed by
the requirements of electroweak symmetry breaking and that the lightest
neutralino be the LSP, whereas the constraint on the relic density is
less severe. At very low $M_{in}$, the electroweak vacuum conditions
become the dominant constraint, and a secondary source of
astrophysical cold dark matter would be necessary to explain the
measured relic density for nearly all values of the soft SUSY-breaking
parameters and $\tbt$. We calculate the neutralino-nucleon cross sections for viable scenarios
and compare them with the present and projected limits from direct
dark matter searches.

\vfill
\leftline{April 2007}
\end{titlepage}

\section{Introduction}

Over the past three and a half decades, the Standard Model (SM) of
particle physics has been remarkably successful at describing the
interactions of elementary particles at or below the weak
scale. However, there are several compelling reasons to expect that
the SM is merely a low-energy effective theory that fits into a larger
framework. Chief among these reasons are the related hierarchy and naturalness problems,
namely the creation and maintenance of a large hierarchy of mass scales despite the fact that
the electroweak Higgs potential is unstable with respect to quantum corrections
within the SM \cite{hierarchy}.
The appearance of supersymmetry (SUSY) at the TeV scale would not only
solve the naturalness problem and  facilitate
the unification of gauge couplings at a high scale as in simple Grand Unified
Theories (GUTs) \cite{gut}, but also predict a light Higgs boson as apparently favoured by the
high-precision electroweak data \cite{erz}. With the additional plausible
assumption of R-parity conservation, the lightest SUSY particle (LSP) is stable and, if
uncharged, is a natural candidate for astrophysical cold dark matter \cite{EHNOS}. For these
reasons, models with SUSY broken at the TeV scale are extensively studied. 

It is evident that SUSY must be broken, since we
have not yet observed any superpartners of SM particles, but the
mechanism of SUSY breaking and how this breaking is communicated to
the observable sector have been the subjects of much
speculation \cite{BIM}. Phenomenologically, the magnitudes of the SUSY-breaking parameters observable at low energies are often calculated by assuming values
of the soft SUSY-breaking parameters at some high input scale and evolving them
down to lower scales using the renormalization-group equations (RGEs) of the
effective low-energy theory. This is generally taken to be the minimal
supersymmetric extension of the SM (MSSM) \cite{mssm}. In the constrained MSSM
(CMSSM)~\cite{cmssm,efgos,efgosi,cmssmnew,cmssmmap,like1,like2}, 
the soft SUSY-breaking parameters are assumed to be universal at
the high scale.  It should be noted, however, that there are many theories of
SUSY breaking in which the soft SUSY-breaking parameters are not universal 
at the input scale~\cite{dterm}.

The CMSSM can be
parametrized at the universality scale by five free input parameters, namely the scalar mass,
$m_0$, the gaugino mass, $m_{1/2}$, the trilinear soft breaking parameter,
$A_0$, the ratio of the Higgs vevs, $\tbt$, and the sign of the Higgs mass parameter, $\mu$. 
The input scale at which universality is assumed in CMSSM models is usually taken to be the
SUSY GUT scale, $M_{GUT} \sim 2 \times 10^{16}$ GeV. However, it may be more
appropriate in some models to assume the soft SUSY-breaking parameters to be
universal at some different input scale, $M_{in}$, which may either be intermediate between
$M_{GUT}$ and the electroweak scale~\cite{EOS06}, the case studied here, or perhaps larger than 
$M_{GUT}$ \cite{pp}.

Specific scenarios in which the soft SUSY-breaking parameters may be
universal at a scale below $M_{GUT}$ occur in models with mixed
modulus-anomaly mediated SUSY breaking (MM-AMSB), also called mirage-mediation~\cite{mixed},
and models with warped extra dimensions~\cite{itoh}.  In the case of mirage-mediation, the
universality scale is the mirage messenger scale, which is predicted
to be $M_{in} \sim 10^{10}-10^{12}$ GeV in the case of KKLT-style moduli
stabilization~\cite{KKLT}. In other models, the universality scale may lie
anywhere between 1 TeV and $M_{Pl}$.

In this paper, we present an in-depth study of the effect on the allowed regions of the CMSSM 
parameter space of lowering the assumed universality scale.
We focus on the dependences of the constraints from cosmology and collider
experiments on the value of $M_{in}$ in such GUT-less scenarios,
paying particular attention to the regions
of parameter space favored by the value of the cold dark matter
relic density inferred from WMAP \cite{WMAP} and other measurements, assuming 
that the cold dark matter is mainly provided by the lightest neutralino $\chi$. 
Within the GUT-less allowed
regions, we also calculate the
neutralino-nucleon cross sections and compare them with present and
expected limits from direct searches for cold dark matter. 

This work is a 
sequel to the exploratory study of GUT-less
CMSSM scenarios 
made in~\cite{EOS06}, in which our attention
was restricted to the case $\tbt = 10$, $A_0 = 0$, $\mu
\ge 0$ and $M_{in} \ge 10^{11.5}$ GeV. We found that, as the universality scale was
reduced to this value, one of the most
dramatic changes was to the footprint in the $(m_{1/2}, m_0)$ plane
of the constraint on the relic abundance of
neutralinos inferred from WMAP {\it et al}. In the standard GUT-scale universality case,
there are three well-defined cosmologically preferred regions of parameter space where
the relic density of neutralinos matches the estimate of the cold dark matter
relic density based on data from WMAP and other observations: the coannihilation region~\cite{stauco},
the rapid-annihilation funnel \cite{efgosi,funnel} and the focus-point region~\cite{focus}. 
In the GUT-less CMSSM scenario \cite{EOS06}, we found
that, as the universality scale is lowered to $M_{in} \sim 10^{12}$
GeV, these regions approach and merge, forming a small WMAP-preferred
island in a sea of parameter space where the neutralino relic density
is too small to provide all the cold dark matter wanted by WMAP. 
We found that, in this case, the only region with a neutralino relic
density that exceeds the WMAP measurement is a `vee' at large $m_{1/2}$, bordering the
region where the stau is the LSP. 

In this paper, we extend the previous analysis to include other values of
$A_0$ for $\tbt = 10$, to the case $\tbt = 50$, and to lower values of $M_{in}$. For this purpose,
we extend the code used previously to evaluate the
cold dark matter density by implementing all coannihilations between the three lightest
neutralinos and the lighter chargino species. As we exhibit explicitly, their inclusion is essential for an accurate calculation of the relic density in some important regions of the GUT-less parameter space.
The second objective of this paper is to calculate the neutralino dark matter scattering
cross sections (both spin-dependent and spin-independent) in such GUT-less models.

The outline of the paper is as follows. In Section 2 we discuss briefly the
renormalizations of the SUSY-breaking contributions to the masses of the squarks, 
sleptons and gauginos as functions of $M_{in}$, as a preliminary to provide background
understanding for some of the results presented later. Then, in Section~3 we
discuss the current experimental, phenomenological and
cosmological constraints on CMSSM scenarios that we use.
Section 4 contains our core discussion of the variation in the allowed
region of parameter space as $M_{in}$ is decreased from the GUT scale down to 
$M_{in} = 10^{9}$ GeV, for both $\tbt = 10$ and $\tbt = 50$. We also
present a separate treatment of the mirage-mediation scenario~\cite{mixed} with
KKLT moduli stabilization~\cite{KKLT}. We then present in Section 5 the corresponding predictions for
neutralino-nucleon scattering cross sections in GUT-less scenarios, and Section~6
summarizes our conclusions. An Appendix
motivates and discusses relevant details of our implementation of multi-channel
neutralino and chargino coannihilation.

\section{Renormalization of SUSY-Breaking Mass Parameters}
\label{sec:renorm}

\begin{figure}
\begin{center}
\mbox{\epsfig{file=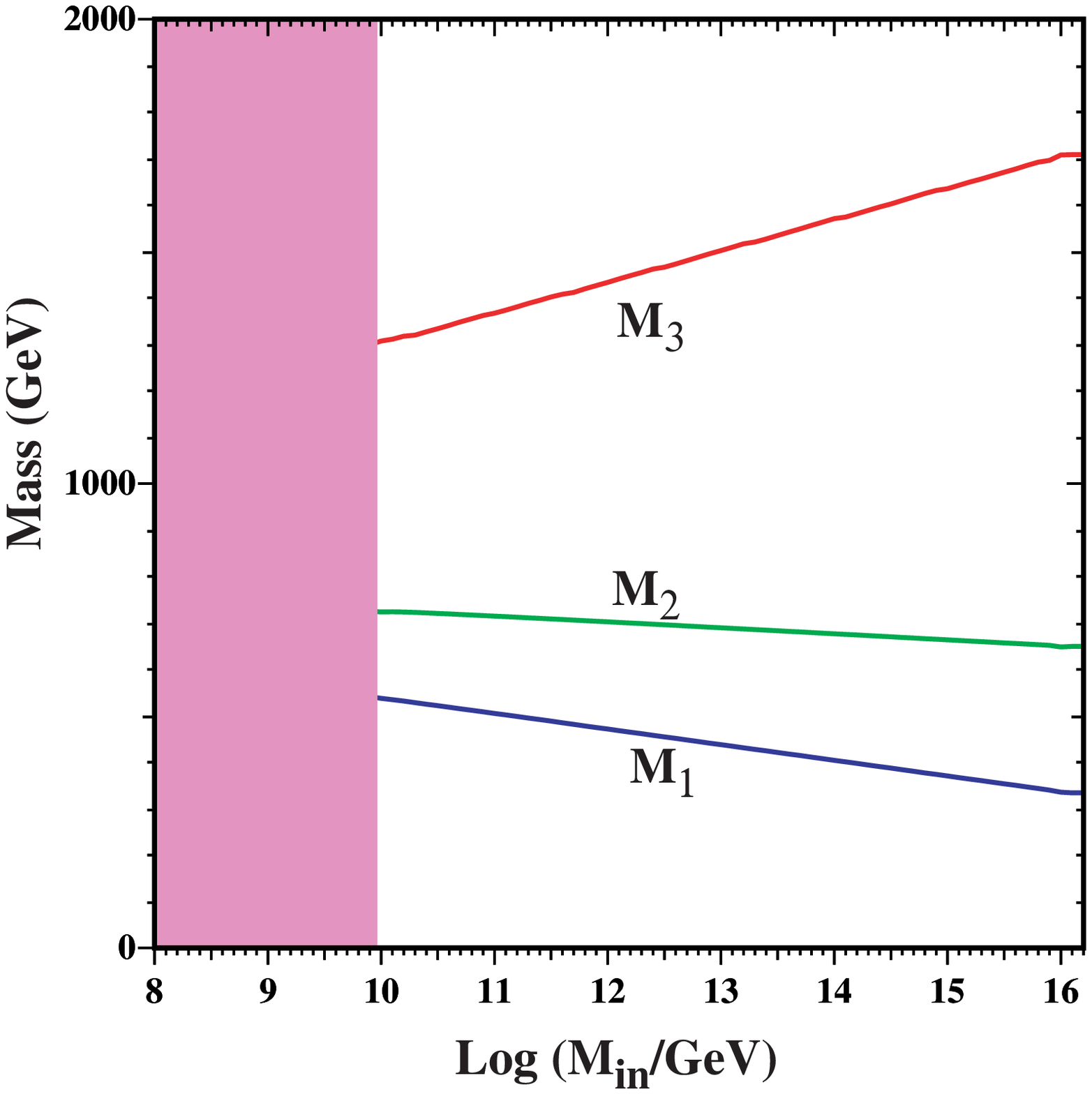,height=7cm}}
\mbox{\epsfig{file=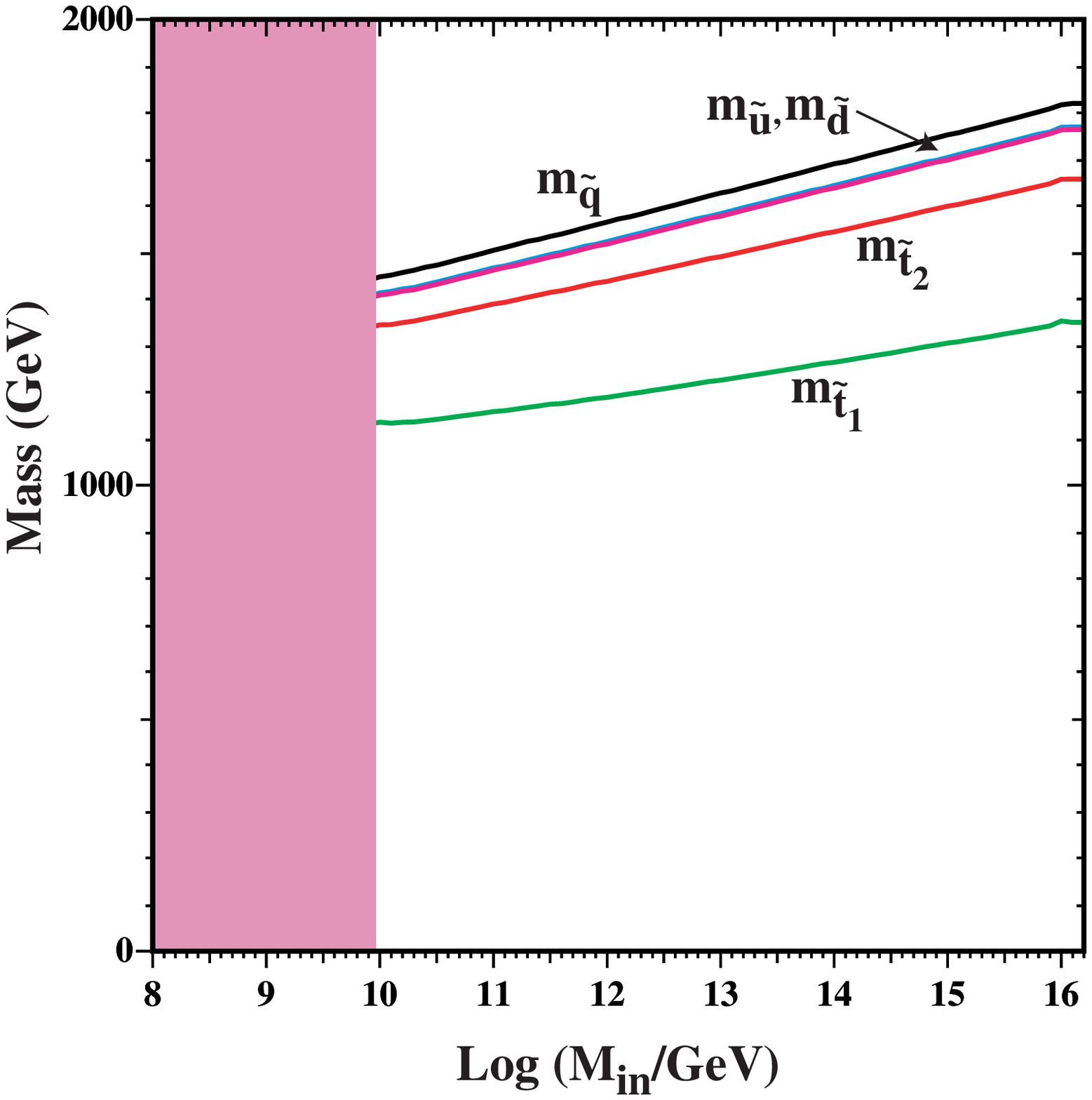,height=7cm}}
\end{center}
\begin{center}
\mbox{\epsfig{file=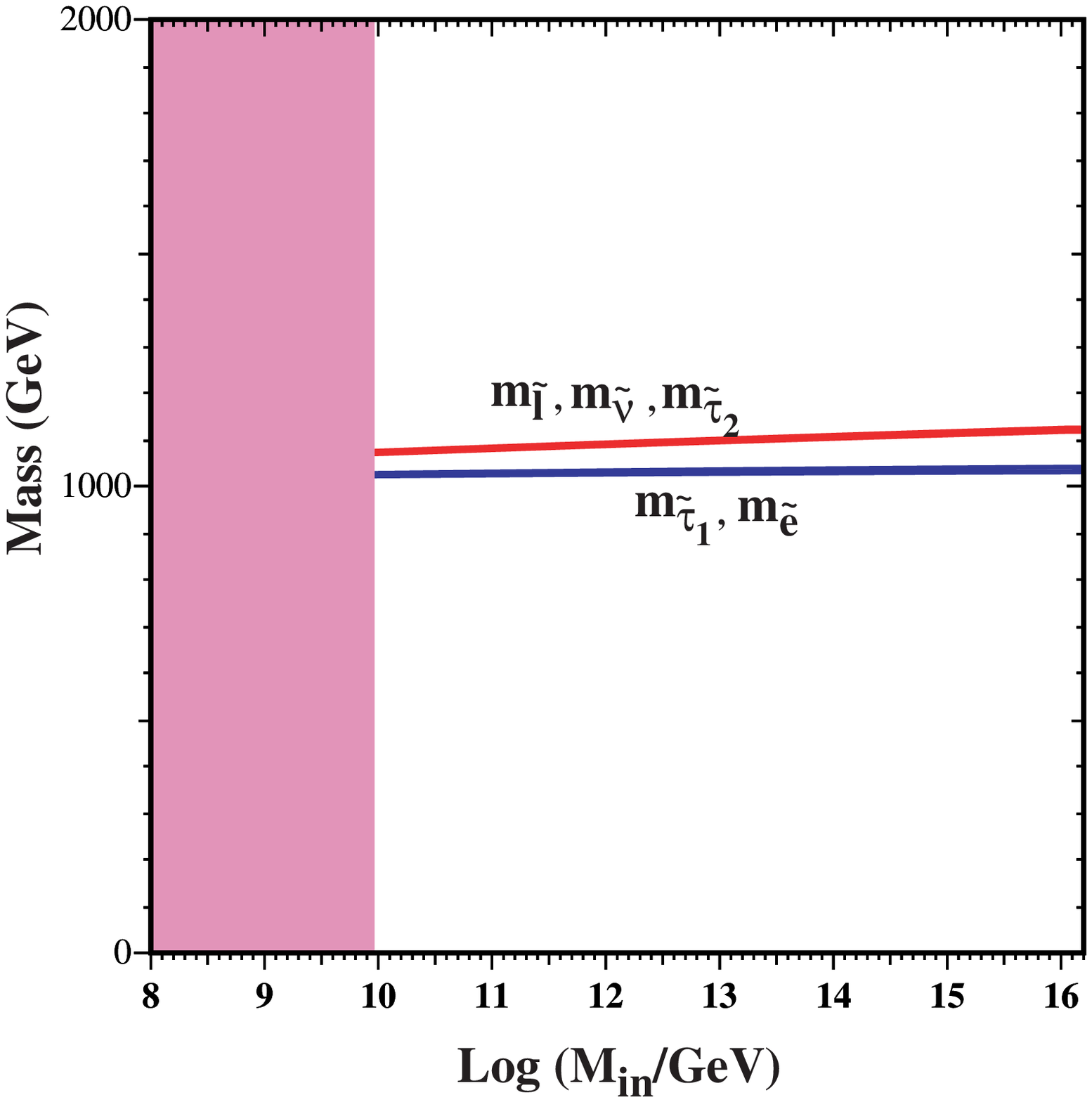,height=7cm}}
\mbox{\epsfig{file=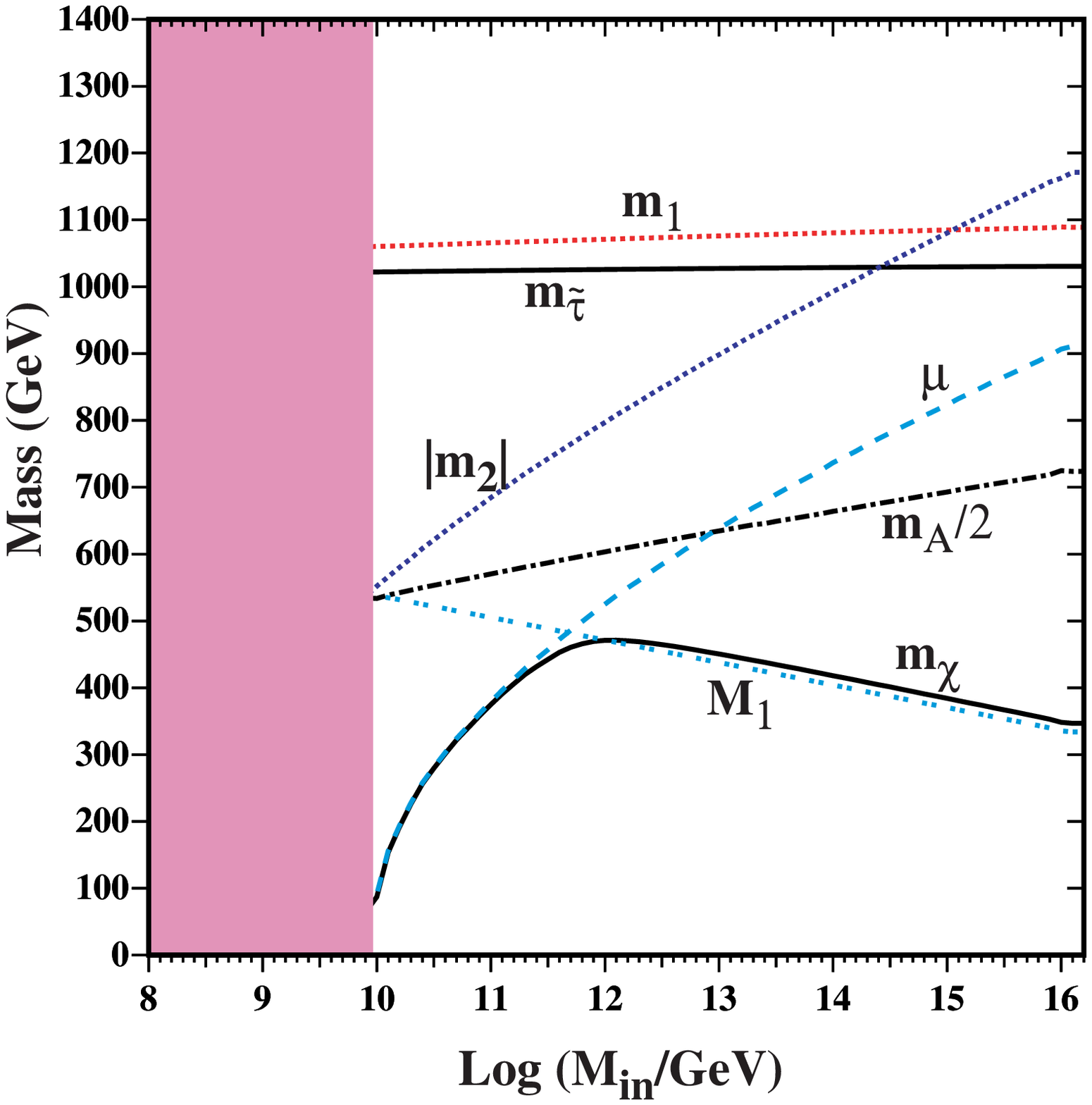,height=7cm}}
\end{center}
\caption{\it
The dependences of observable sparticle mass parameters on the input scale
$M_{in}$ at which they are assumed to be universal: (a) gaugino masses
$M_{1,2,3}$, (b) squark
masses, (c) slepton mases, and (d) Higgs ($m_{1,2}, m_A$), stau and the lightest
neutralino $\chi$ masses, as well as $\mu$ and the U(1) gaugino mass $M_1$.
The calculations are made for the representative case $m_{1/2} = 800$~GeV,
$m_0 = 1000$~GeV, $A_0 = 0$, $\tan \beta = 10$ and $\mu > 0$.}
\label{fig:masses}
\end{figure}

In order to understand the changes in the allowed
regions in the $(m_{1/2},m_0)$ plane of the CMSSM that occur
as $M_{in}$ is lowered, it is necessary first to understand the consequences 
for the observable sparticle masses of lowering the
universality scale. In the CMSSM with universality
imposed at the GUT scale, the one-loop renormalizations of the
gaugino masses $M_a$, where $a=1,2,3$, are the same as those for the
corresponding gauge couplings, $\alpha_a$. Thus, 
at the one-loop level the gaugino masses at
any scale $Q \leq M_{GUT}$ can be expressed as 
\beq
M_a(Q) = \frac{\alpha_a(Q)}{\alpha_a(M_{GUT})}M_a(M_{GUT}),
\eeq
where $M_a(M_{GUT}) = m_{1/2}$. On the other hand, in a
GUT-less CMSSM, where the gauge-coupling strengths run at all scales
below the GUT scale but the soft SUSY-breaking parameters run only below
the lower universality scale, $M_{in}$, at which all the gaugino masses are
assumed to be equal to $m_{1/2} = M_a(M_{in})$, we have
\beq
M_a(Q) = \frac{\alpha_a(Q)}{\alpha_a(M_{in})}m_{1/2}
\label{gaugino}
\eeq
at the one-loop level.
Since the runnings of the coupling strengths in GUT and GUT-less CMSSM
scenarios are identical, the low-energy effective soft gaugino
masses, $M_a(Q)$, in GUT-less cases are less separated and closer to
$m_{1/2}$ than in the usual GUT CMSSM, as seen explicitly in panel (a)
of Fig.~\ref{fig:masses}~\footnote{Note that in making this plot we have included the full
two-loop renormalization-group equations for the gaugino masses, which are
not identical to those for the gauge couplings, although the difference is not
very striking.}.

The soft SUSY-breaking scalar masses are renormalized by both
gauge and (particularly in the cases of third-generation sfermions)
Yukawa interactions, so the running is somewhat more
complicated. At the one-loop level one can summarize the effects of renormalizations at any
$Q \leq M_{in}$ as
\beq
m_{0_i}^2(Q) = m_0^2(M_{in})+C_i(Q,M_{in})m_{1/2}^2,
\label{scalar}
\eeq
where we make the CMSSM assumption that the $m_0^2(M_{in})$ are
universal at $M_{in}$, and the $C_i(Q,M_{in})$ are renormalization coefficients that
vanish as $Q \goto M_{in}$. We display in panels (b) and (c) of
Fig.~\ref{fig:masses} the two-loop-renormalized soft SUSY-breaking masses of
the the first- and second-generation left- and right-handed squarks,
${\tilde q_{L,R}}$, the stop mass eigenstates, ${\tilde t_{1,2}}$, and the
left- and right-handed sleptons, ${\tilde \ell_{L,R}}$.
We see again that in GUT-less cases the soft SUSY-breaking scalar masses
are less separated and closer to $m_0$ than in the usual GUT-scale CMSSM.

In the CMSSM, the electroweak vacuum conditions are used to fix the
values of $|\mu|$ and $m_A$. Although we use the full two-loop renormalizations,
insight into the effects of varying $M_{in}$ on the required values of $|\mu|$ and
$m_A$ can be gleaned from simple leading-order expressions. 
The tree-level solution for $\mu$ is
\beq
\mu^2 = \frac{m_1^2-m_2^2\ttbt}{\ttbt-1} - \frac{M_Z^2}{2},
\label{eq:mu}
\eeq
where $m_1$ and $m_2$ are the soft Higgs masses associated with $H_1$
and $H_2$, respectively. The variation of $\mu$ with $M_{in}$ for one fixed pair of values of
$(m_{1/2}, m_0)$ is seen in panel (d) of Fig.~\ref{fig:masses}, where we see that the
solution of (\ref{eq:mu}) for $\mu^2$ becomes negative and unphysical for $M_{in} < 10^{10}$~GeV.
For this value of $M_{in}$, values of $m_0 > 1000$~GeV would not yield physical
electroweak vacua. One can see from (\ref{scalar}) and panels (b)
and (c) of Fig.~\ref{fig:masses} that, as $M_{in}$
decreases, the soft scalar masses remain closer to the input
value, $m_0$. This has the converse result that, for any fixed $m_{1/2}$, as
the universality scale is lowered, $\mu^2$ changes sign 
and becomes unphysical at smaller values of $m_0$, causing the 
upper boundary of the unphysical region to creep down
farther into the $(m_{1/2},m_0)$ plane.
This explains the encroachment of the upper-left excluded regions in the
$(m_{1/2},m_0)$ planes shown later in Figures \ref{fig:mint} - \ref{fig:mint502}, 
as $M_{in}$ decreases.

The weak-scale value of $m_A$ decreases with $M_{in}$ logarithmically,
as also seen in panel (d) of Fig.~\ref{fig:masses}, 
and also in panels (c) and (d) of Fig. 3 of~\cite{EOS06}. In addition to its importance for the
direct detection of the near-degenerate $A, H$ and $H^\pm$ bosons, this
feature is important indirectly for several aspects of our later discussion. One is the 
constraints from heavy-flavour physics to be discussed in the next section: since 
$b \goto s \gamma$ and $\bmm$ at large $\tbt$ have important contributions from
the exchanges of heavier Higgs bosons, the impact of these constraints increases
as $m_A$ decreases and hence as $M_{in}$ decreases. A second impact of
$m_A$ is on the cold
dark matter density: since a rapid-annihilation funnel appears when $m_\chi \simeq m_A/2$,
for fixed values of the other parameters such as $\tbt, m_0$ and $A_0$, this funnel
appears at lower $m_\chi$ and hence $m_{1/2}$ as $M_{in}$ decreases. Finally,
another potential impact is on the spin-independent neutralino dark-matter scattering cross section, 
which receives a significant contribution from heavy Higgs exchange,
as discussed later.

In addition to the excluded regions in the upper left corners of each of
the $(m_{1/2},m_0)$ planes shown in Figures
\ref{fig:mint}-\ref{fig:mint502} where electroweak symmetry breaking
is not obtained, we see a second major excluded region in the lower
right corner of each panel. In these regions of the plane, the
lightest stau, $\stau_1$, becomes lighter than the lightest
neutralino, resulting in a charged LSP, which is incompatible with
general arguments from astrophysics and cosmology. As we see from
(\ref{scalar}), as $M_{in}$ decreases the positive coefficient
$C_{\stau_1}$ also decreases because $M_{in}$ is approaching the low scale,
$Q$.  Hence $m_{\stau_1}$ gets progressively closer to $m_0$ for any fixed
$m_{1/2}$, as seen in panel (c) of Fig.~\ref{fig:masses}. At the same time, the 
gaugino masses remain closer to
$m_{1/2}$ as $M_{in}$ decreases, implying that, as long as the lightest
neutralino remains essentially a bino, its mass becomes a larger portion of the universal 
gaugino mass.  This can be seen in panel (d)
of Fig.~\ref{fig:masses}, where for this particular point in the
$(m_{1/2},m_0)$ plane, the LSP mass tracks that of the bino for
$M_{in} \gtrsim 10^{12}$ GeV. As a result, for fixed $m_{1/2}$ and $m_0$, as
the universality scale $M_{in}$ is lowered from $M_{GUT}$, initially $m_{\chi_1}$ increases and
$m_{\stau_1}$ decreases.  Hence, as $M_{in}$ decreases for
any fixed $m_{1/2}$, a larger value of $m_0$
is required to enforce the condition $m_{\chi_1} \leq m_{\stau_1}$. For this reason,
the lower-right excluded regions in the
$(m_{1/2},m_0)$ planes shown in Figures \ref{fig:mint} - \ref{fig:mint502}
initially expand as $M_{in}$ decreases.

 However, since $|\mu|$ decreases as
$M_{in}$ decreases, as discussed above, below a certain value of $M_{in}$, 
$|\mu|$ becomes small
enough that the lighter Higgsino takes over as the LSP, with a mass
that decreases as $|\mu|$ continues to decrease.  In panel (d) of Fig.~\ref{fig:masses},
one can see that, for $M_{in} \lesssim 10^{11}$ GeV, the LSP is
sufficiently Higgsino-like that its mass is nearly identical to
$|\mu|$. Since the boundary of the disallowed stau LSP region is determined by equality between the masses of the stau and
the lightest neutralino, this boundary therefore
falls to lower $m_0$ when $M_{in}$ is decreased
below the bino-Higgsino cross-over point, as is seen in in the $(m_{1/2},m_0)$ planes shown 
later in Figures \ref{fig:mint} - \ref{fig:mint502}.

\section{Experimental, Phenomenological and Cosmological Constraints}

Our treatments of experimental, phenomenological and cosmological
constraints essentially follow those in~\cite{EOS06}, but with differences that we describe
below.

\subsection{LEP Experimental Constraints}\label{sec:LEPconstraints}

The appropriate LEP lower
limit on the chargino mass for the class of CMSSM models discussed here is
$m_{\chi^\pm} > 104$~GeV\cite{LEPsusy}, and the 
nominal effective lower limit on the mass of the
lightest Higgs boson $h$ is 114~GeV~\footnote{We implement this constraint by 
calculating the lightest Higgs mass with 
the previous version of the {\tt FEYNHIGGS} code \cite{FeynHiggs}, which
incorporates a direct interface with the underlying CMSSM parameters, and
allowing a possible error of 1.5~GeV to account for possible higher-order contributions. 
We have verified that
the numerical difference from the more recent version of {\tt FEYNHIGGS} is
considerably smaller than our error allowance.}~\cite{LEPHiggs}.
However, in addition to displaying
the direct position of the 114 GeV bound in the GUT-less parameter space, 
we also calculate and display the 95\% CL limit obtained by
combining the experimental likelihood, ${\cal L}_{exp}$, from 
direct searches at LEP~2 and a global
electroweak fit, convolved with the theoretical and parametric errors in $m_h$ \footnote{
We thank A.~Read for providing the LEP $CL_s$ values.}, which provides a more exact
(and relaxed) interpretation of the LEP Higgs limit within the MSSM. The top mass used in these 
calculations is $m_t = 171.4 \pm 2.1$ GeV \cite{mt}.

We note that
one can use (\ref{scalar}) to predict how the impact of the LEP Higgs mass constraint 
varies with $M_{in}$. We recall that the mass of the lightest scalar MSSM Higgs boson
$m_h < M_Z$ at tree level, but is renormalized by an amount that depends logarithmically on
$m_{\stop}$. Eq.~(\ref{scalar}) shows that $m_{\stop}$ decreases as $M_{in}$ is lowered for 
any fixed $m_{1/2}$ and $m_0$. However, $m_{\stop}$ also increases with $m_{1/2}$. 
Thus, one should expect that the LEP Higgs constraint moves to
larger $m_{1/2}$ as the universality scale is lowered.

\subsection{Muon Anomalous Magnetic Moment}

It is well known that the measurement by the BNL g-2 Collaboration~\cite{g-2} disagrees 
significantly with the Standard Model if $e^+ e^-$ annihilation data are used to calculate 
the Standard Model contribution, but there is no significant discrepancy if this is calculated 
using $\tau$-decay data~\cite{Davier}. In view of the lack of consensus on the interpretation 
of the measurement of $a_\mu = (g_{\mu}-2)/2$, we use it only as part of our motivation
for  restricting our study to the
case $\mu > 0$. However, if the $e^+ e^-$ estimate of the hadronic
contribution to the Standard Model calculation is accepted, one finds~\cite{ICHEP}:
\begin{eqnarray}
a_\mu ({\rm theory}) & = & (11659180.5 \pm 5.6) \times 10^{-10}, \\
a_\mu ({\rm experiment}) & = & (11659208.0 \pm 6.3) \times 10^{-10},
\end{eqnarray}
yielding a discrepancy \cite{Davier}
\begin{equation}
\Delta a_\mu \; = \; (27.5 \pm 8.4)  \times 10^{-10},
\label{amu}
\end{equation}
which would be a 3.3-$\sigma$ effect. In the plots discussed later, we display
the corresponding 2-$\sigma$ range, namely
\begin{equation}
10.7 \times 10^{-10} \; < \; \Delta a_\mu \; < \; 44.3 \times 10^{-10}.
\label{g-2}
\end{equation}

\subsection{B Decay Observables}

We consider two constraints provided by B decay: one is the agreement between
experiment and theory for $b \goto s \gamma$ \cite{bsgth}, and the other is the experimental
upper limit on $\bmm$ decay. The recent measurements of $B^\pm \goto \tau^\pm \nu$
decay do not yet impinge significantly on the parameter space we explore in this paper.

In the case of $b \goto s \gamma$, we use the estimate $BR(b \to s \gamma)
= (3.15 \pm 0.23) \times 10^{-4}$~\cite{Misiak} for the SM contribution at 
NNLO~\footnote{We note that the 
dominant theoretical error
due to the renormalization-scale uncertainty it is not Gaussian, and hence
we add it linearly rather than in quadrature with the other errors.},
and the code of Gambino and Ganis~\footnote{We thank Geri Ganis for
a recent update to this code.}  to
calculate the MSSM contribution to the decay amplitude at NLO in QCD.
As for the present
experimental rate for $b \goto s \gamma$ decay, we use the range
\begin{equation}
BR(b \to s \gamma) = (3.55 \pm 0.24^{+0.12}_{-0.13} ) \times 10^{-4}
\label{hfag}
\end{equation}
as recommended by the HFAG \cite{bsg,hfag}.
The first of the errors in (\ref{hfag}) is the combined statistical and systematic experimental
error. The second set of errors result from theoretical uncertainties and corrections. 
These are combined linearly with the scale uncertainty in the calculation.
We recall that $b \to s \gamma$ joins $a_\mu$ in disfavouring $\mu < 0$.

In the case of $\bmm$ decay, we calculate the rate in the MSSM using~\cite{bmm,eosp},
and we use the experimental upper limit
\begin{equation}
BR(\bmm) < 1.0 \times 10^{-7}
\end{equation}
reported by CDF~\cite{CDFbmumu}. We also display in Figures \ref{fig:mint50} - \ref{fig:mint502}
projected future sensitivities of the Tevatron and LHC experiments (a factor of 5 times
lower than the current limit). As already noted, the
the impact of the $\bmm$ constraint is important at large $\tbt$, and increases
as $m_A$ decreases and hence as $M_{in}$ decreases.

\subsection{Neutralino Relic Density}

As aleady mentioned, 
we assume that the neutralino LSP constitutes essentially all the cold dark matter,
for which we consider the allowed range to be~ \cite{WMAP}:
\begin{equation}
0.0855 < \Omega_\chi h^2 < 0.1189,
\label{WMAP}
\end{equation}
as mandated by WMAP and other observations.

As discussed in more detail in the Appendix,
we have included in our calculation of the neutralino relic density, for the first time, 
all the processes for coannihilation between
the three lightest neutralino states $\chi_{1,2,3}$, as well as with the lighter chargino $\chi^\pm$
and with sleptons. The importance of $\chi_1 - \chi_2 - \chi^\pm$ coannihilation has long
been recognized within the context of the GUT-scale CMSSM \cite{co1,efgos}. Near the top-left
boundary of the allowed region in the $(m_{1/2}, m_0)$ plane, the lightest neutralino is
Higgsino-like, near the bottom of the allowed region the lightest neutralino is bino-like,
and the bino and Higgsino masses cross over along some intermediate contour. Near this cross-over
line, and particularly where it intersects the left boundary of the allowed region in 
the $(m_{1/2}, m_0)$ plane, $\chi_1 - \chi_2 - \chi^\pm$ coannihilation is important in the
GUT-scale CMSSM.

In the GUT-less CMSSM, as we show later, there are interesting regions of the
$(m_{1/2}, m_0)$ plane at small $M_{in}$ where the $\chi_3$ mass comes within
${\cal O}(200)$~GeV of the $\chi_2$ mass, and coannihilation processes involving the $\chi_3$
can no longer be neglected. The reason for this, despite the relatively large $\chi_3 - \chi_2$
mass difference, is that the couplings of the Higgsino-like $\chi_3$ to relevant final states are
significantly larger than the corresponding $\chi_2$
couplings. Regions of the plane where $\chi_2$ and $\chi_3$ are
degenerate are present at most values of $M_{in}$, though they
typically occur when $\chi_1$ is much lighter than the other
neutalinos. For low $M_{in}$, however, there is in fact a
near-degeneracy of all three of the lightest neutralinos as well as
the lighter chargino. It is therefore necessary to include all
coannihilations involving the three lightest neutralinos and the
lighter chargino, as detailed in the Appendix.

In addition, we implement here various improvements to our previous treatment 
of the dark-matter density in regions where
rapid annihilation via a direct-channel Higgs pole is important. Specifically, we have included
further crossed-channel contributions to $WW$, $ZZ$ and less important processes.

\section{Evolving Impact of the Cold Dark Matter Constraint}

We now discuss the evolution of the dark matter
constraint as the scale at which the soft supersymmetry breaking
parameters are universal is lowered from the GUT scale. 
We assume $m_{t} = 171.4$ GeV in this analysis. Deviation by a few
GeV from this value would result in some change to the exact positions
and shapes of the regions preferred by WMAP, but our results are
quite general. We recall that, as usual in the CMSSM, the value of the Higgs
mixing parameter $\mu$ is fixed by the electroweak vacuum conditions, leaving
its sign as a free parameter. Motivated by $a_\mu$ and $b \to s \gamma$,
we consider only $\mu > 0$, though a similar
analysis could be carried out for negative $\mu$.  In
Sections \ref{sec:lowtanb} and \ref{sec:hightanb} we discuss in detail
the effects of lowering the universality scale for two values of the
ratio of the Higgs vevs, $\tbt = 10$ and $\tbt = 50$. We take $A_0 = 0$
throughout Sections \ref{sec:lowtanb} and \ref{sec:hightanb}, and
examine the impact of deviation from this assumption in Section \ref{sec:A0}.
Related mirage-mediation models are discussed in Section~\ref{sec:mirage}.

\subsection{Low $\tbt$
\label{sec:lowtanb}}

\begin{figure}
\begin{center}
\mbox{\epsfig{file=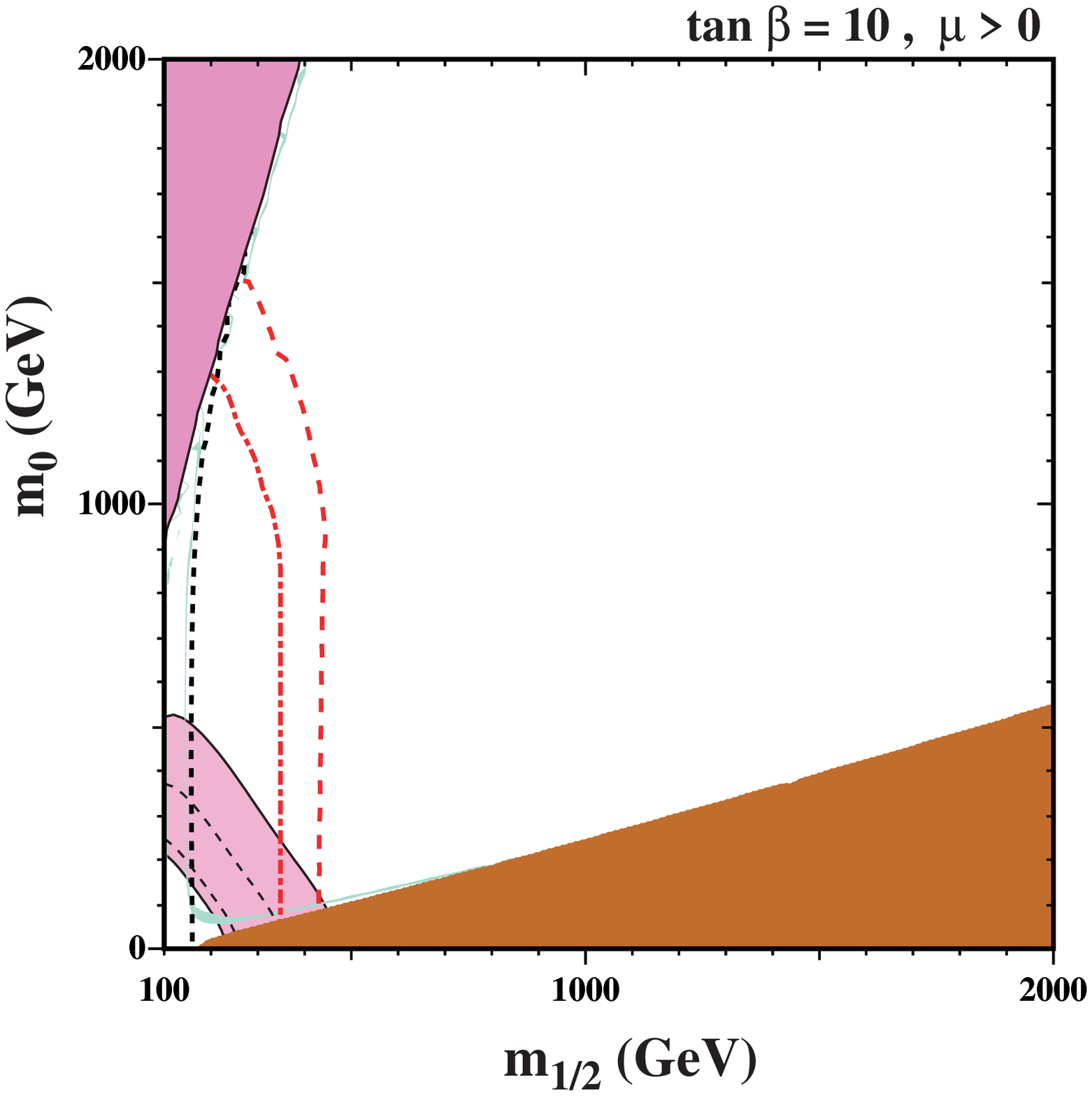,height=7cm}}
\mbox{\epsfig{file=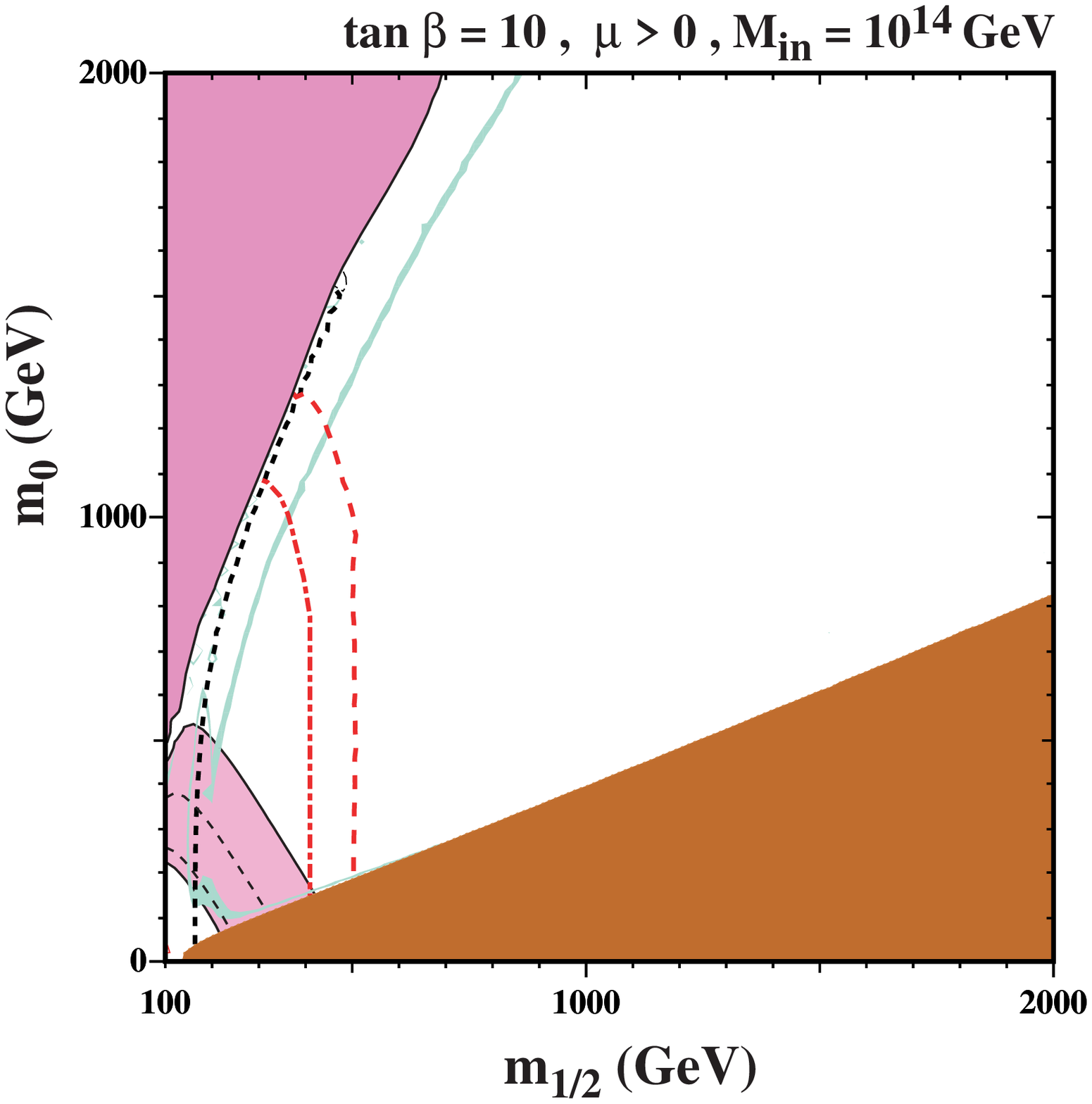,height=7cm}}
\end{center}
\begin{center}
\mbox{\epsfig{file=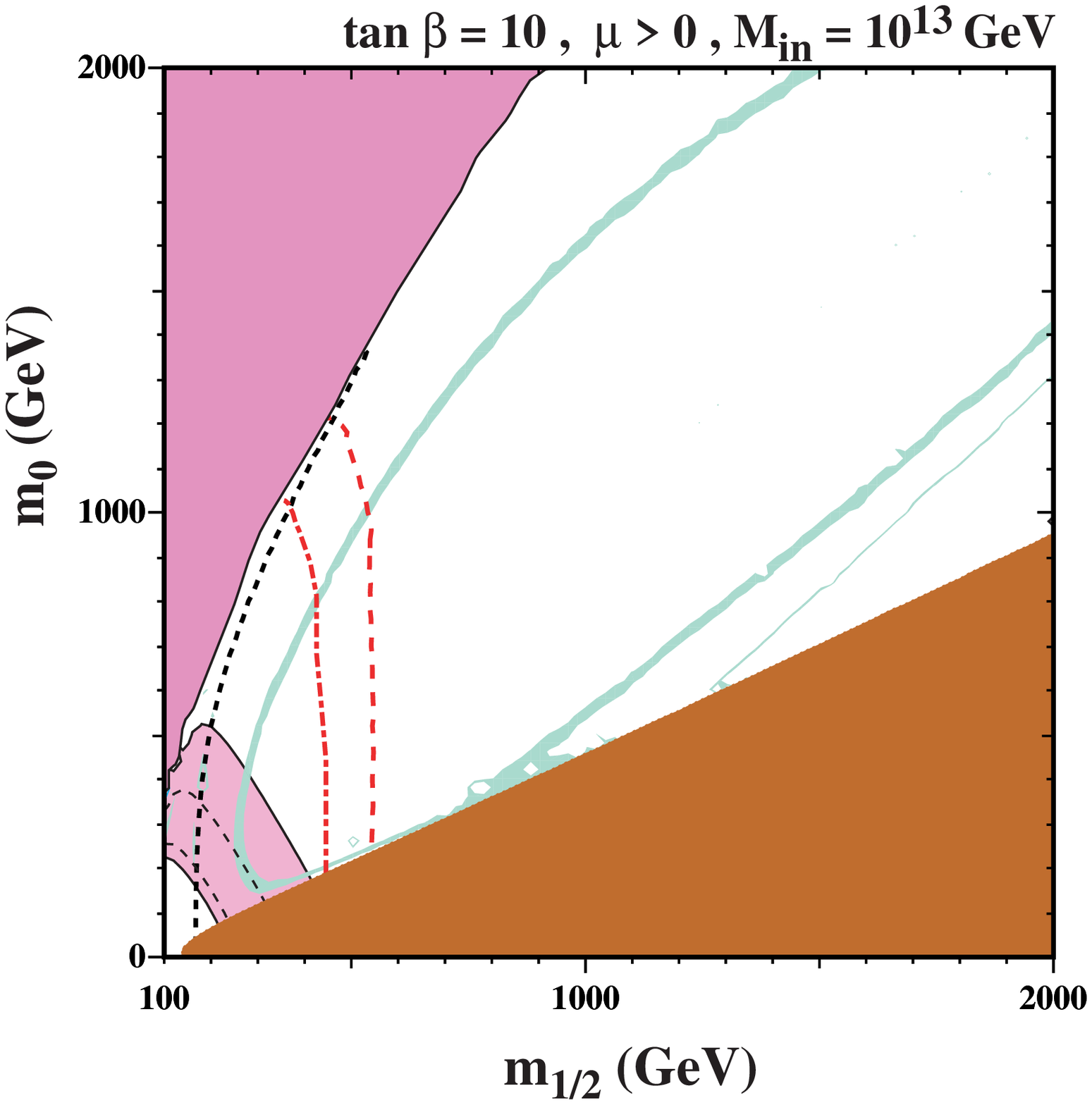,height=7cm}}
\mbox{\epsfig{file=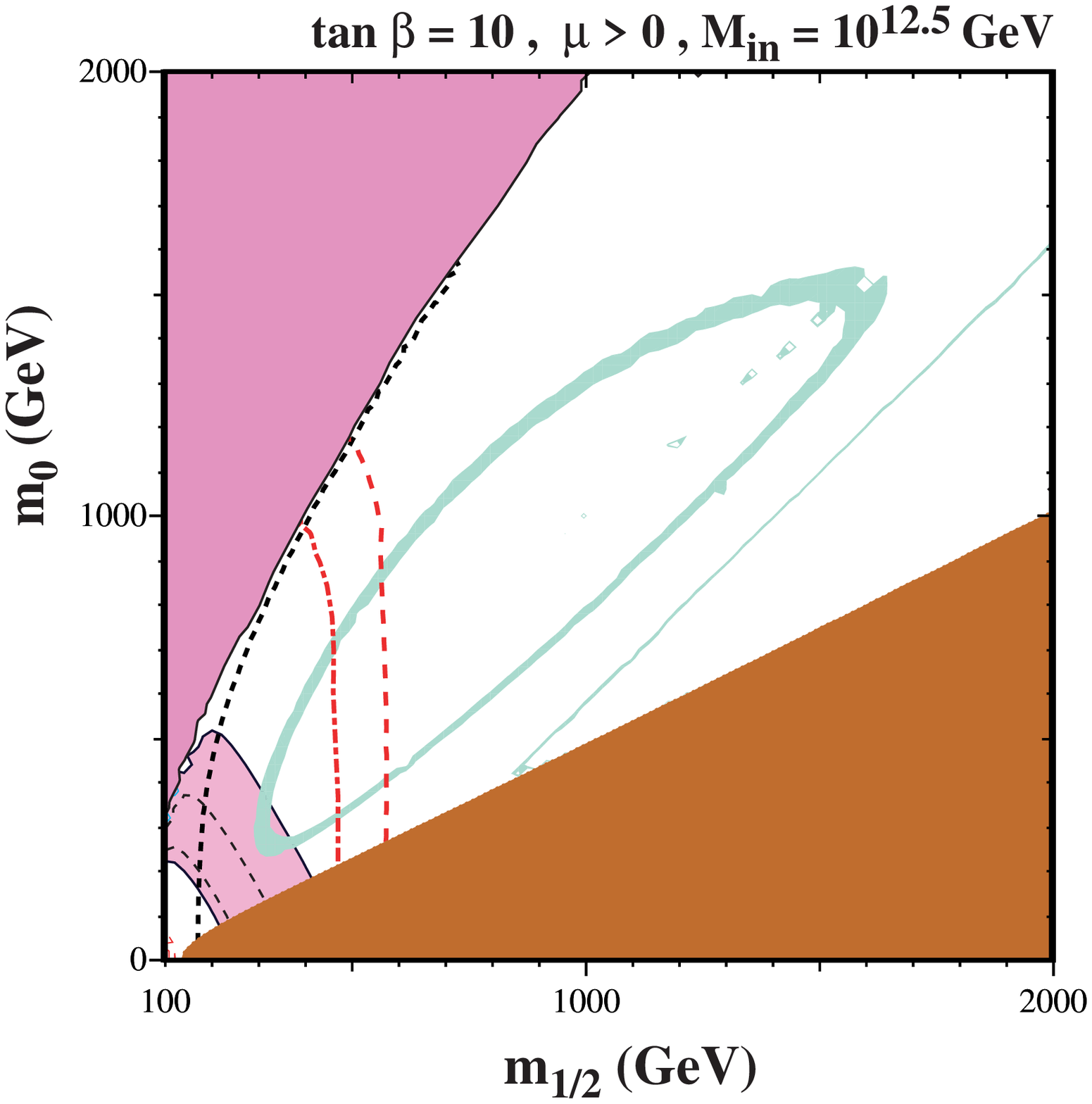,height=7cm}}
\end{center}
\caption{\it
Examples of $(m_{1/2}, m_0)$ planes with $\tan \beta = 10$ and 
$A_0 = 0$ but with  different values of $M_{in}$.
(a) The CMSSM case with $M_{in} = M_{GUT} \sim 2 \times 10^{16}$~GeV, 
(b) $M_{in} = 10^{14}$ GeV, 
(c) $M_{in} = 10^{13}$ GeV and (d) $M_{in} = 10^{12.5}$ GeV. 
In each panel, we show contours representing the LEP lower limits on the
chargino mass (black dashed line), a Higgs mass of 114 GeV (red dashed),
and the more exact (and relaxed) Higgs bound (red dot-dashed). We also show the region
ruled out because the LSP would be charged (dark red shading), and
that excluded by the electroweak vacuum condition (dark pink shading). The region favoured 
by the WMAP range $\Omega_{CDM} h^2 =
0.1045^{+0.0072}_{-0.0095}$ has light turquoise shading, and the region 
suggested by $g_\mu - 2$ at 2-$\sigma$ has medium (pink) shading, with
the 1-$\sigma$ contours shown as black dashed lines.}
\label{fig:mint}
\end{figure}

\begin{figure}
\begin{center}
\mbox{\epsfig{file=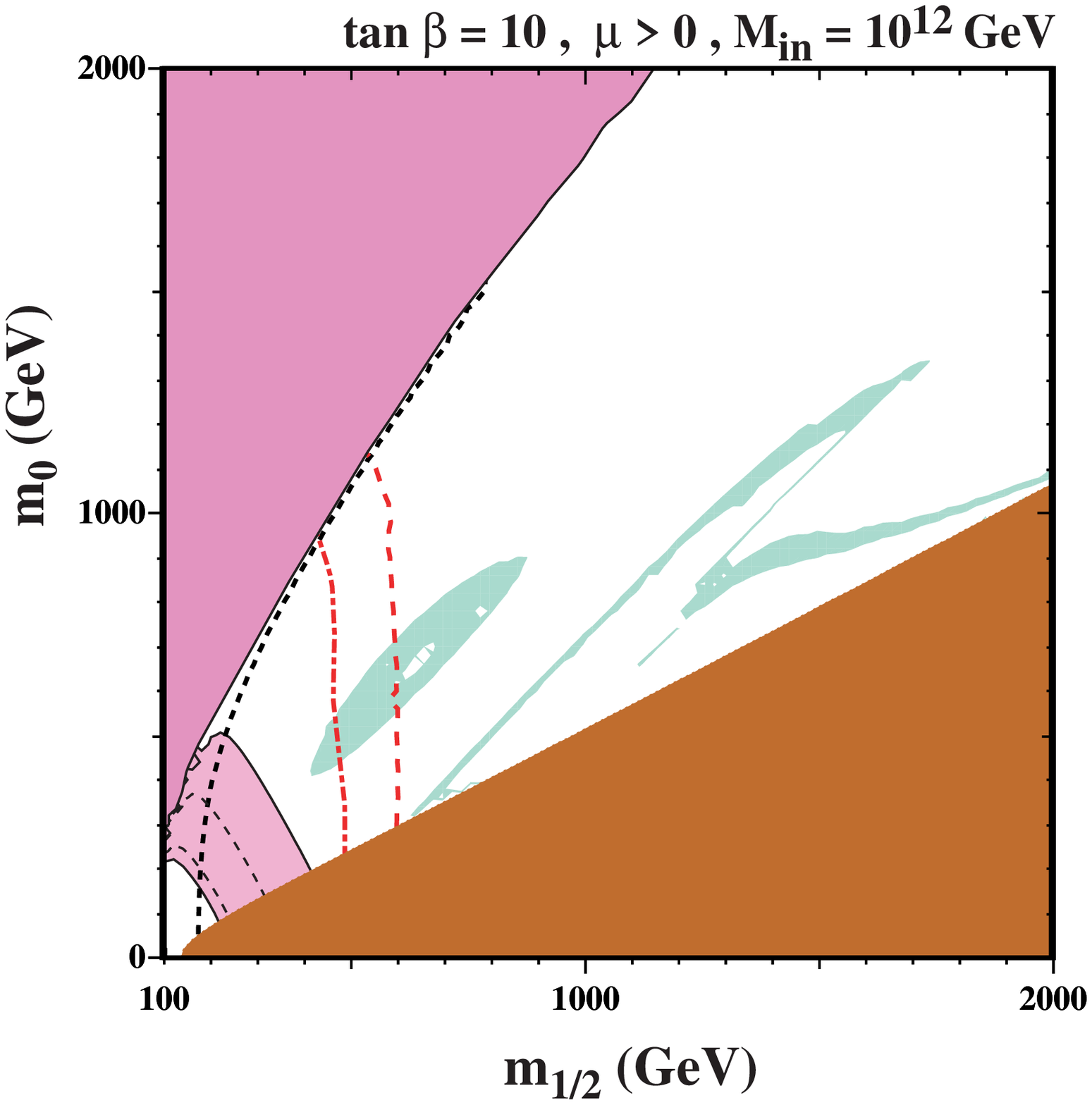,height=7cm}}
\mbox{\epsfig{file=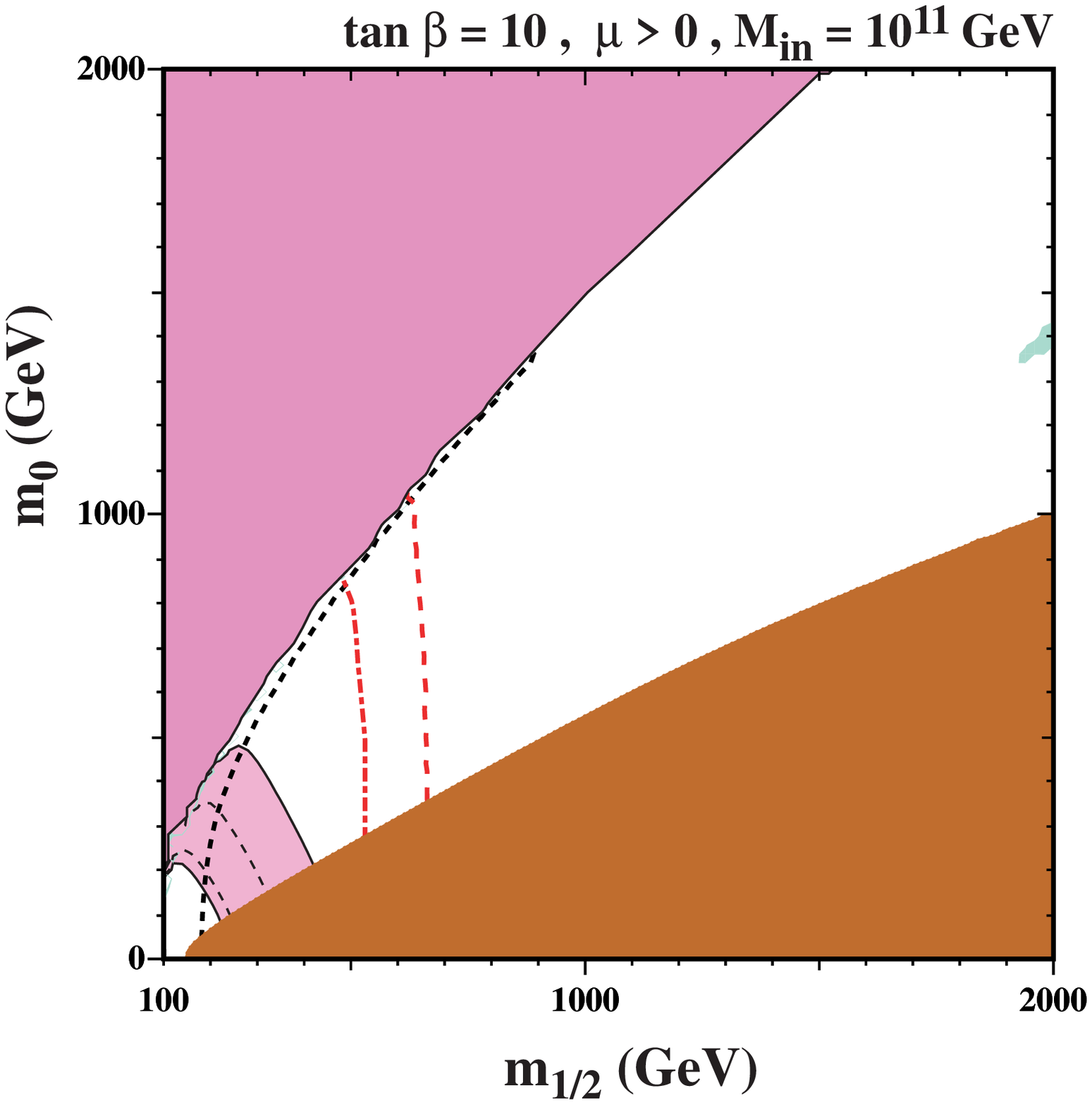,height=7cm}}
\end{center}
\begin{center}
\mbox{\epsfig{file=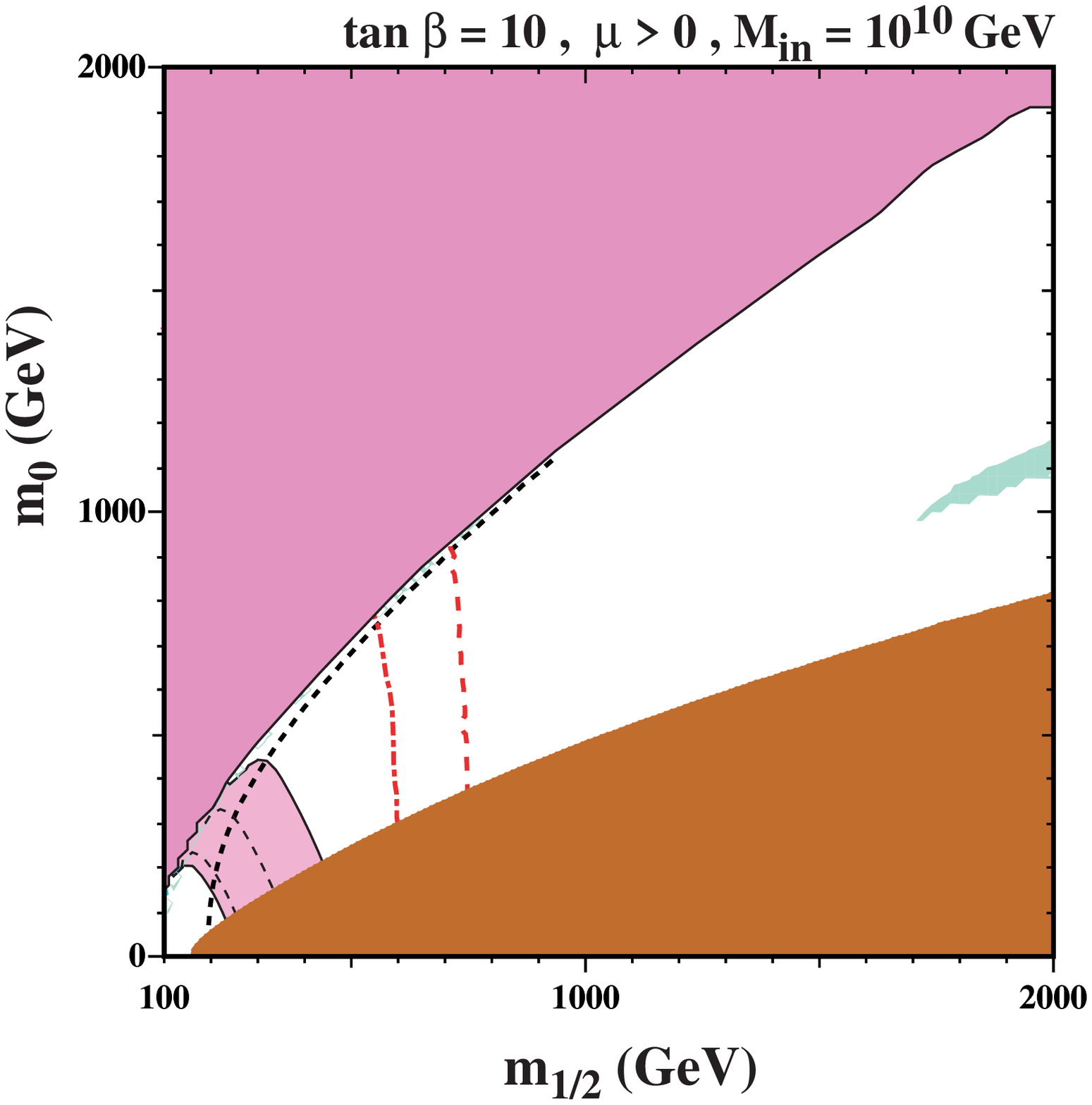,height=7cm}}
\mbox{\epsfig{file=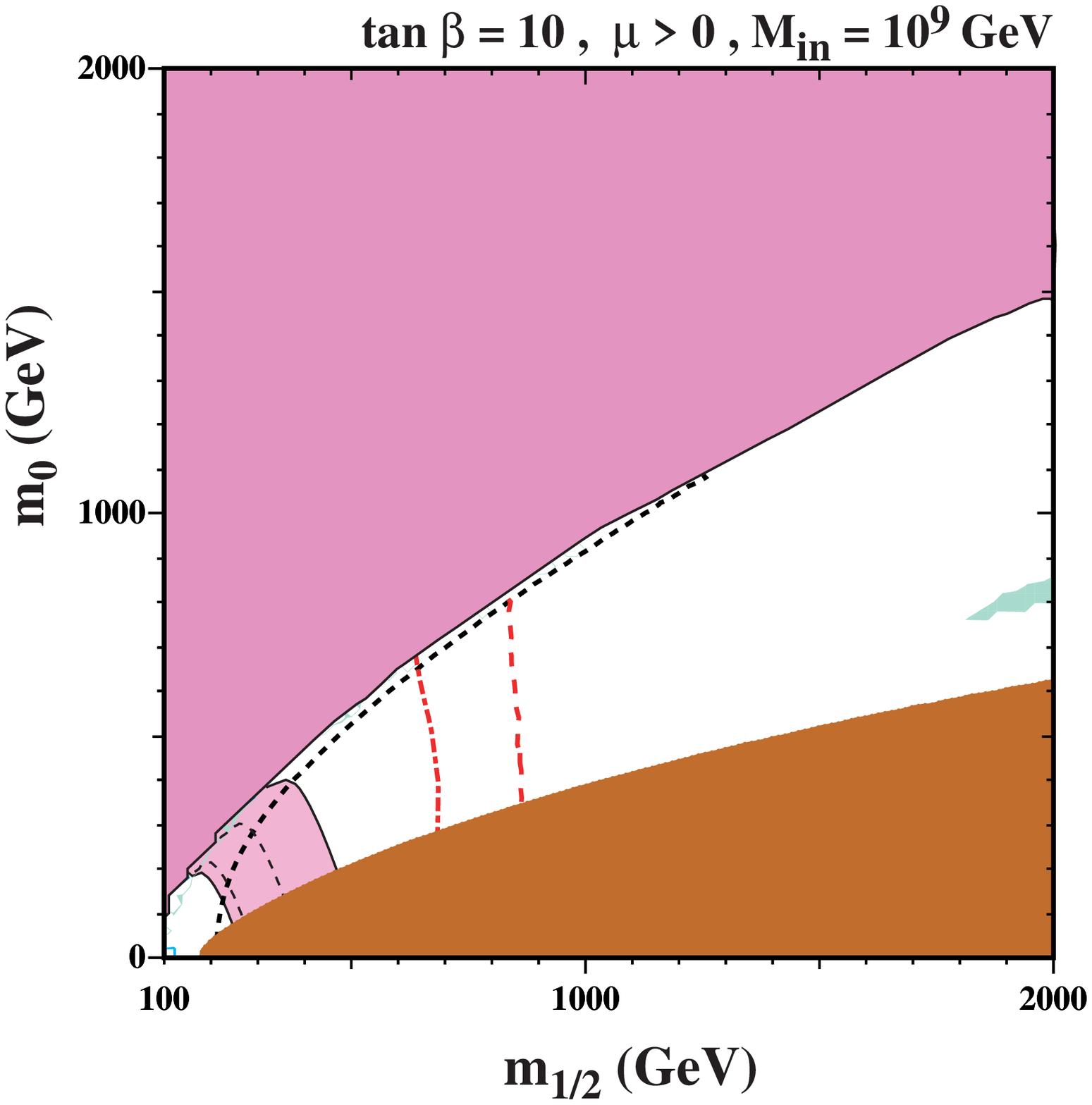,height=7cm}}
\end{center}
\caption{\it
Further examples of $(m_{1/2}, m_0)$ planes with $\tan \beta = 10$ and 
$A_0 = 0$ but with  different values of $M_{in}$:
(a) $M_{in} = 10^{12}$ GeV, 
(b) $M_{in} = 10^{11}$ GeV, 
(c) $M_{in} = 10^{10}$ GeV and (d) $M_{in} = 10^{9}$ GeV.
The various contours and shadings are the same as for Fig.~2.}
\label{fig:mint2}
\end{figure}

The evolution of the WMAP-preferred region in the $(m_{1/2},m_0)$
plane as the universality scale
is lowered has been discussed previously in~\cite{EOS06} for $\tbt = 10$ and $M_{in}
\geq 10^{11.5}$ GeV. The WMAP-preferred regions
found in this analysis, along with constraints from colliders, are shown in Figs.~\ref{fig:mint} and
\ref{fig:mint2} for several values of $M_{in}$. To begin, we look
first at the usual GUT-scale CMSSM scenario, shown in 
panel (a) of Fig.~\ref{fig:mint}. One can see the $\chi
- \stau$ coannihilation region bordering the excluded stau LSP region
for $330 \lesssim m_{1/2} \lesssim 900$ GeV. Values of $m_{1/2}$ below
this range are excluded by the LEP Higgs constraint.  Near $m_{1/2} =
900$ GeV, the coannihilation strip dips down into the region where the
$\stau$ is the LSP. The focus point appears as a very thin strip
tracking the border of the region excluded by the electroweak symmetry
breaking condition at $m_0 > 1500$ GeV. The LEP chargino
bound also follows this boundary. The rapid-annihilation
funnel is not present at $\tbt = 10$ for $M_{in}$ at the GUT scale,
but will appear as the universality scale is lowered and also at
larger $\tbt$.

As found in~\cite{EOS06}, there are already changes as the universality scale
is lowered to $M_{in} = 10^{14}$ GeV, shown in Panel (b) of Figure~\ref{fig:mint}. 
The allowed focus-point region starts to separate from the LEP
chargino bound, moving to larger $m_{1/2}$. Notice also that this
strip does not join smoothly with the coannihilation strip, but
instead is deflected due to rapid $h$ annihilation near $m_{1/2} \sim 150$
GeV. The region where the relic density falls in the WMAP range is thereby
pushed inside the LEP chargino mass bound.  However, this behavior
occurs at low values of $m_{1/2}$ which are excluded by the LEP Higgs
bound as well.

For $M_{in} = 10^{13}$ GeV, shown in panel (c) of Fig.~\ref{fig:mint}, we notice
that, as foreseen in Section~\ref{sec:renorm}, the regions excluded by the electroweak vacuum conditions and because the
stau would be the LSP are encroaching further into the plane as $M_{in}$
decreases, and the LEP Higgs bound is moving to larger $m_{1/2}$.
We see in panel (c) of Fig.~\ref{fig:mint}
that the allowed focus-point region also dips
further down, away from the electroweak vacuum condition boundary, while the
coannihilation strip moves up and farther away from the region where the
stau is the LSP. In fact, the focus-point and coannihilation regions connect, forming an
slender atoll extending to $(m_{1/2}, m_0) \sim (2850, 2400)$~GeV
(beyond the displayed region of the plane), inside which the relic 
density of neutralinos is too
large. Another remarkable feature at this value of
$M_{in}$ is the appearance of the rapid-annihilation funnel, familiar 
in the GUT-scale CMSSM at
large $\tan \beta$, but an unfamiliar feature for $\tbt = 10$.  In the
narrow space between the underside of the atoll and the thin 
WMAP-preferred strip lying approximately $100-200$ GeV below it, $2m_{\chi} \sim m_A$ and
direct-channel annihilation processes are
enhanced, causing the relic density to drop below the value determined by
WMAP.

As the universality scale is further decreased to $M_{in} = 10^{12.5}$~GeV,
as shown in panel (d) of Fig.~\ref{fig:mint}, 
the atoll formed by the conjunction of what had been the focus-point and
coannihilation strips has shrunk, so that it lies entirely within the range of
$(m_{1/2}, m_0)$ shown in panel (d)\footnote{We note a string of bubbles
intruding into the atoll, which are due to a significant enhancement of
$t-$channel exchange in $\chi_2 \chi_2 \goto h+(H,A)$. The analysis of these
possible regions of small relic density would require a complete treatment of
poles, including finite-width effects, which we do not attempt here.}. 
We now see clearly two distinct
regions of the plane excluded due to an excess relic density of
neutralinos; the area enclosed by the atoll and the slice between the
lower funnel wall and the boundary of the already-excluded $\stau$-LSP
region.

The four panels of Figure \ref{fig:mint2} show the consequences of
lowering the universality scale even further, down as far as $M_{in} =
10^{9}$~GeV. In panel (a) for $M_{in} = 10^{12}$~GeV, 
the focus-point and coannihilation regions
are fully combined and the atoll has mostly filled in to become a small island of acceptable relic
density.  To the right of this island is a strip that is provided by the
lower funnel wall.  The strip curves slightly as
$m_{1/2}$ increases then takes a sharp plunge back down towards the
boundary of the region where the stau is the LSP, a feature
associated with the $\chi \chi \goto h + A$ threshold. Reduction in
the universality scale from this point results in the lower funnel wall being pushed down into the
excluded $\stau$ LSP region and total evaporation of
the island.

As the universality scale decreases further in panels (b), (c) and (d)
for $M_{in} = 10^{11}$~GeV, $10^{10}$~GeV and $10^9$~GeV, 
respectively, we see only a
small residual turquoise region at large $m_{1/2}$ where
the relic density is within the WMAP limits.  At all other points in the visible part
of the $(m_{1/2},m_0)$ plane the relic density of neutralinos is too low
to provide fully the cold dark matter density preferred by WMAP {\it et al}. 
Of course, these SUSY models
would not be excluded if there is another source of cold dark matter in the
universe.

In these last four panels, we notice that the boundary of the region
where the stau is the LSP is retreating back down to smaller
$m_0$, as expected from the discussion of evolution with $M_{in}$ of the 
masses of the stau and the lightest neutralino given in Section~\ref{sec:renorm}.

\subsection{High $\tbt$
\label{sec:hightanb}}

\begin{figure}
\begin{center}
\mbox{\epsfig{file=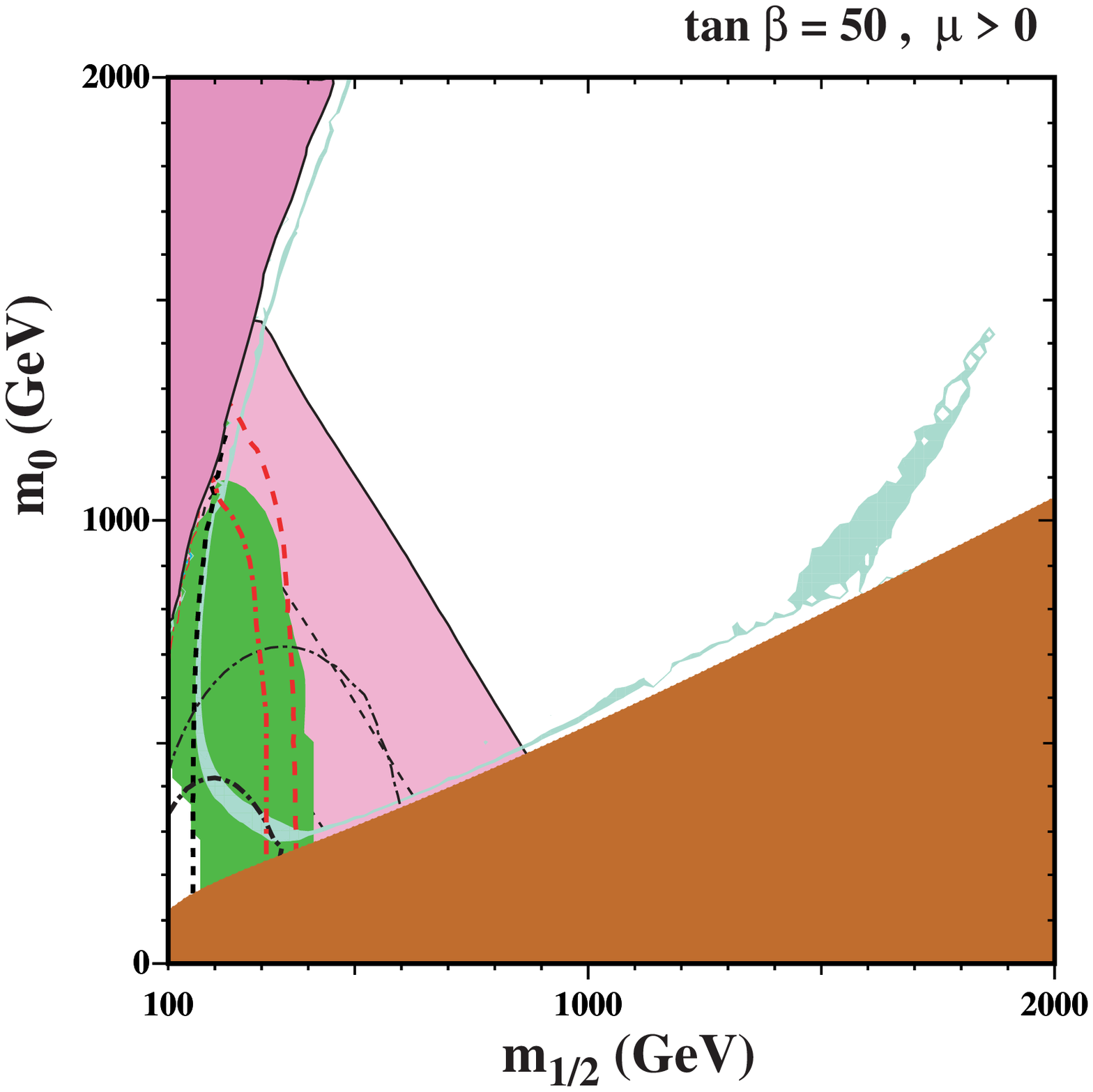,height=7cm}}
\mbox{\epsfig{file=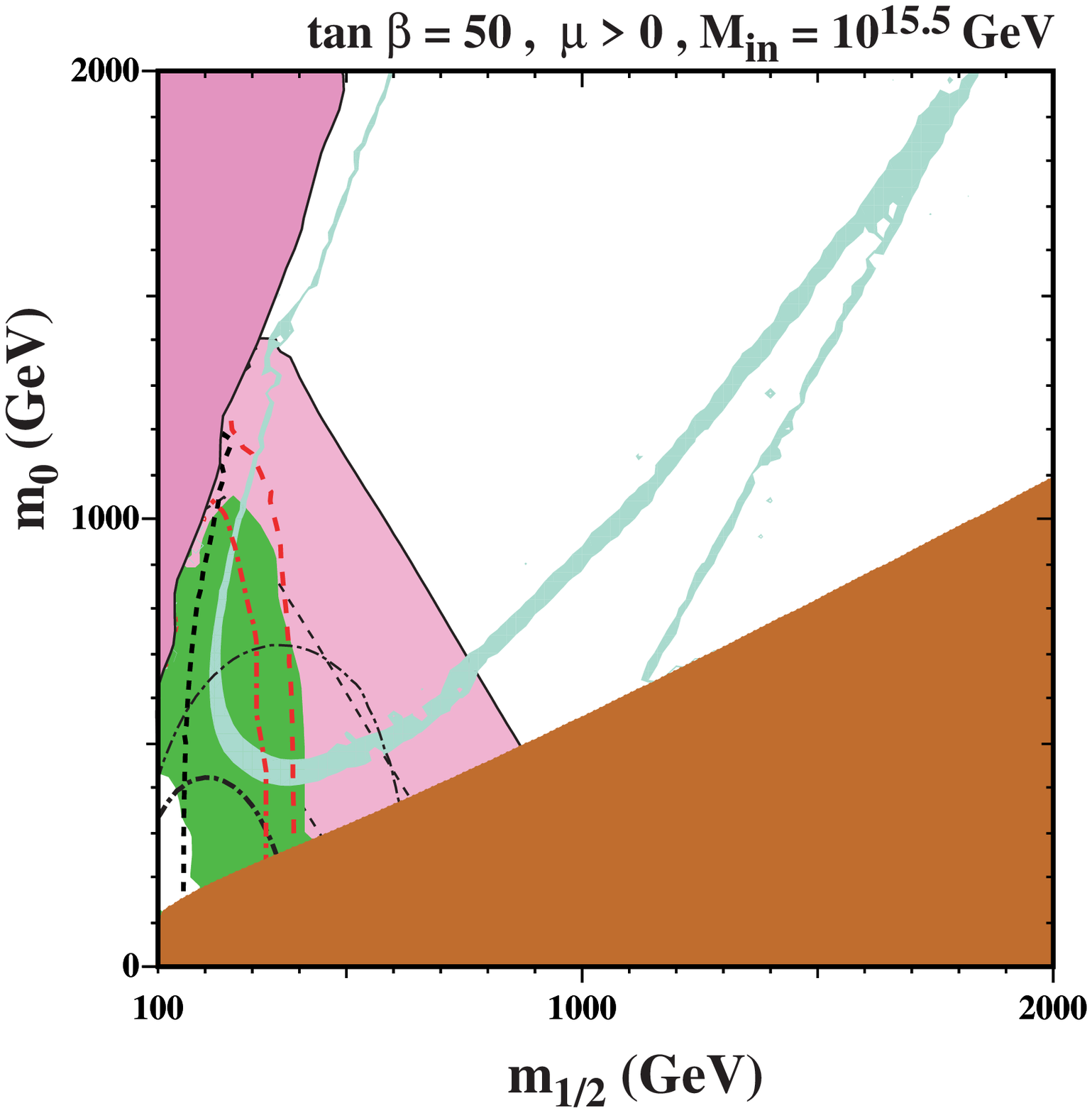,height=7cm}}
\end{center}
\begin{center}
\mbox{\epsfig{file=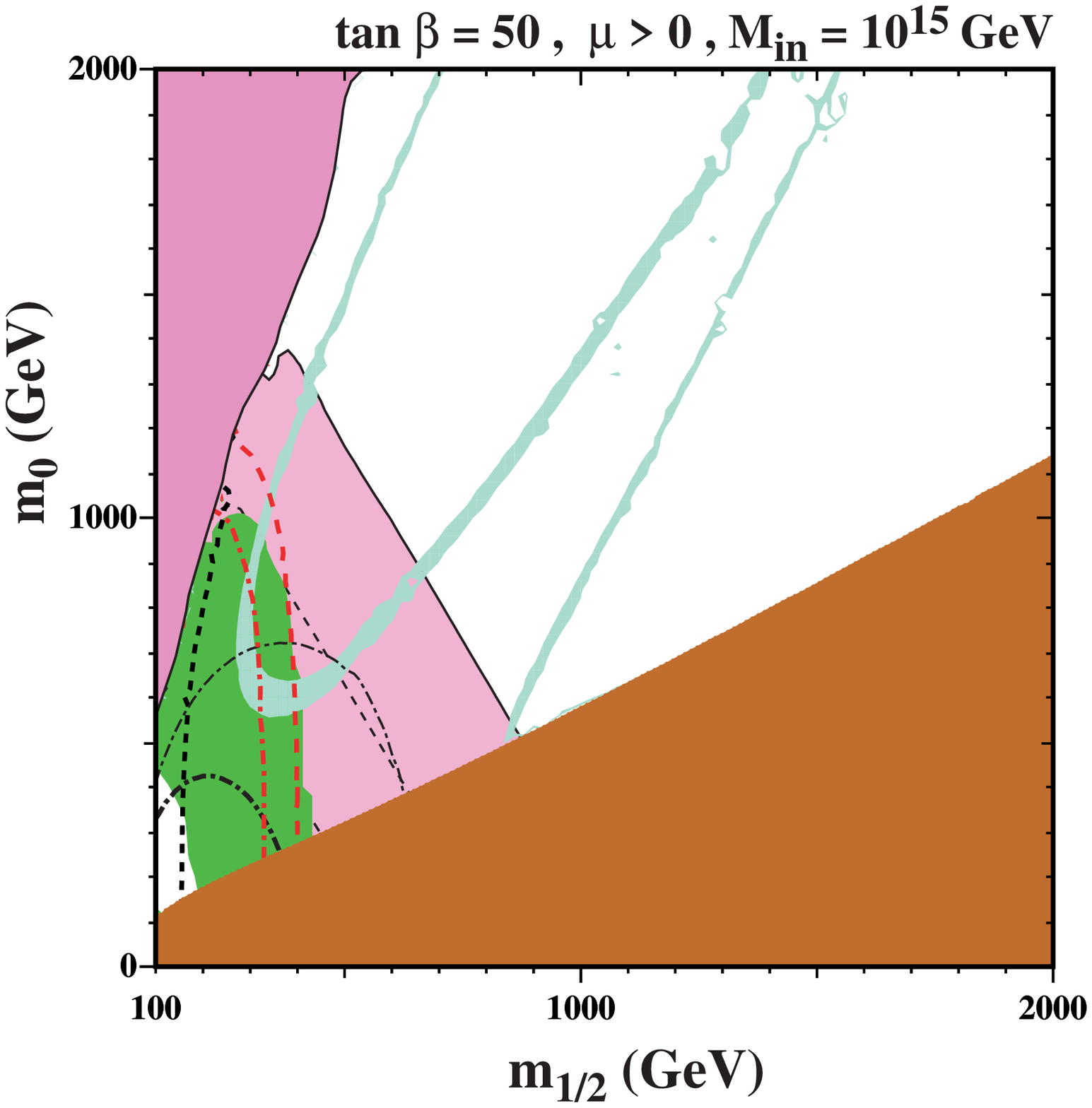,height=7cm}}
\mbox{\epsfig{file=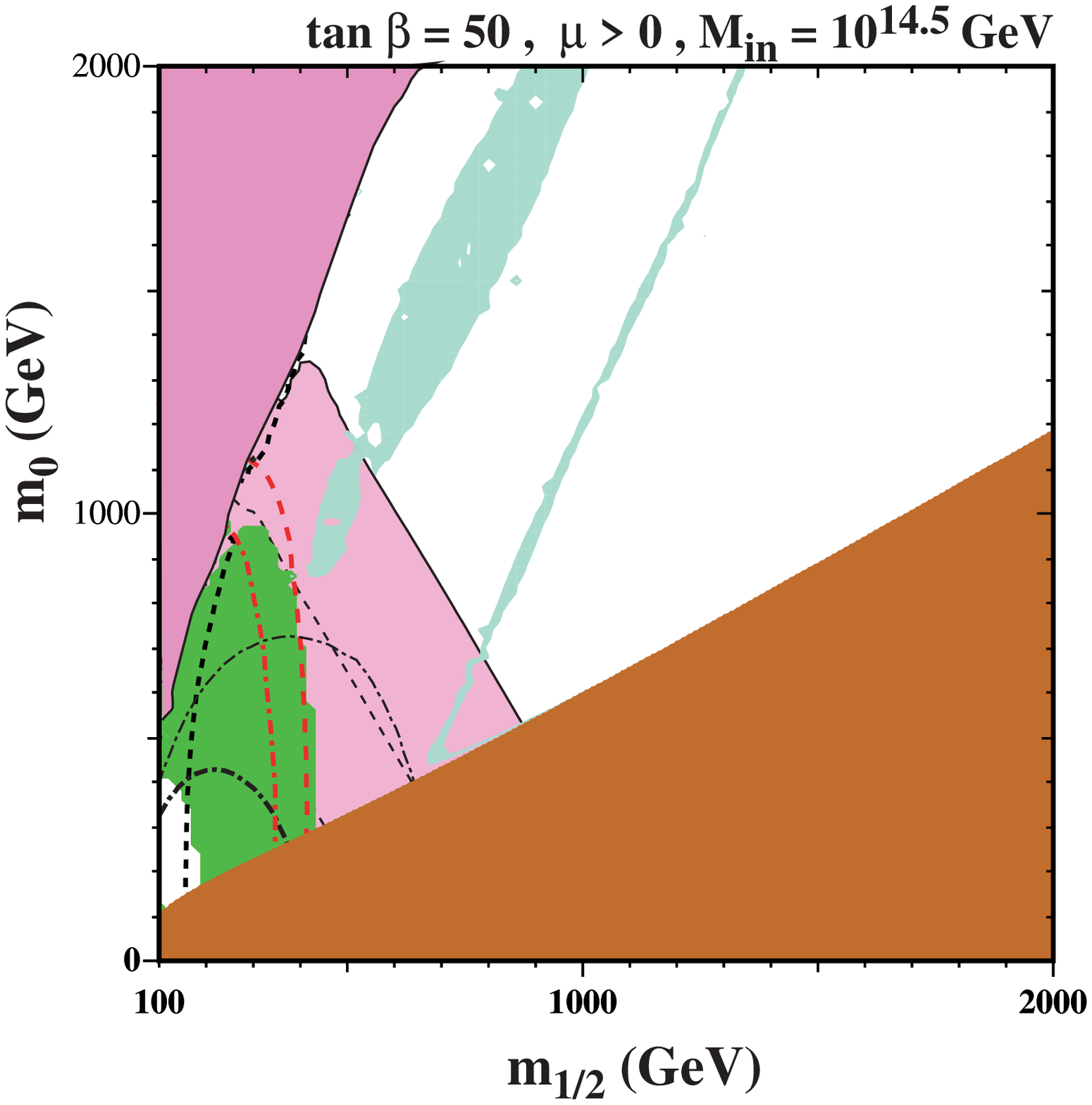,height=7cm}}
\end{center}
\caption{\it
Examples of $(m_{1/2}, m_0)$ planes with $\tan \beta = 50$ and 
$A_0 = 0$ but with  different values of $M_{in}$.
(a) The CMSSM case with $M_{in} = M_{GUT} \sim 2 \times 10^{16}$~GeV, 
(b) $M_{in} = 10^{15.5}$ GeV, 
(c) $M_{in} = 10^{15}$ GeV and (d) $M_{in} = 10^{14.5}$ GeV. In
addition to the constraints enumerated in the caption to
Fig.~\ref{fig:mint}, we also show the regions ruled out by $b
\to s \gamma$ decay~\protect\cite{bsg,hfag,bsgth} (medium green
shading) and black dot-dashed contours representing the current CDF
limit on the rate of $\bmm$($1 \times 10^{-7}$) and a projected sensitivity of the
Tevatron and the LHC experiments ($2 \times 10^{-8}$).
}
\label{fig:mint50}
\end{figure}

The situation at larger $\tbt$ looks somewhat different at first
glance. In the GUT-scale
CMSSM case, shown in panel (a) of Fig.~\ref{fig:mint50}, 
we see the familiar regions excluded because of
a $\stau$ LSP and the electroweak vacuum
condition. The LEP Higgs and chargino mass bounds have impacts
similar to those in the low-$\tbt$ scenario. The
region excluded by $b\goto s \gamma$ decay has grown substantially,
and a new region excluded by the limit $BR(\bmm) > 1 \times 10^{-7}$ appears
at low $(m_{1/2}, m_0)$, which is, however, already excluded by other constraints.
As for the relic density, the focus-point region is visible as a strip tracking
the electroweak vacuum condition for $m_0 > 1050$ GeV, whereas the
region preferred by WMAP is excluded by the LEP Higgs
constraint at smaller $m_0$.
Along the excluded $\stau$ LSP boundary, we see that the familiar
coannihilation strip is truncated at low $m_{1/2}$ by the Higgs and chargino mass
constraints, and also by $\bmm$. Following this strip to larger $m_{1/2}$, 
there is the familiar rapid-annihilation funnel, where 
$2m_{\chi_1} \sim m_A$ and the relic density is 
kept in the range preferred by WMAP by annihilations through the
direct-channel $A$ and $H$ poles, which lifts away from the excluded region. 

However, at large $\tbt$, even small changes in the universality scale make a
dramatic difference in the appearance of the regions preferred by WMAP.
At $M_{in} = 10^{15.5}$~GeV, as seen in panel (b) of Fig.~\ref{fig:mint50},
the coannihilation strip and rapid-annihilation funnel have joined to create a large funnel region that
extends to $(m_{1/2}, m_0) \sim (1850, 2000)$~GeV. Inside the funnel boundary, effects
such as rapid annihilation near the $A$ pole and the coannihilations of neutralinos with light 
sleptons combine to cause the relic
density to fall below the range preferred by WMAP. In this
region of low relic density, the lightest neutralino is bino-like, and the dominant annihilations are into
$b\bar{b}$ and $\tau \bar{\tau}$ pairs. As in the GUT-scale
universality scenario, the focus-point region is cut off at $m_0 \sim 950$
GeV by the LEP Higgs constraint. At this universality
scale, values of $m_{1/2} > 600$~GeV are compatible also with $m_0 > 2000$~GeV, beyond the
displayed region of the $(m_{1/2}, m_0)$ plane.

As the universality scale is further reduced to $M_{in} = 10^{15}$ GeV,
we see in panel (c) of Fig.~\ref{fig:mint50} that the funnel is elongated further and opens 
wider at the top, while simultaneously the focus-point region falls significantly below the zone 
excluded by the electroweak
vacuum conditions. In addition, the bulk region, where the upper
funnel wall connects to the focus point, has now shifted
to larger $m_{1/2}$, so that it lies mostly outside the LEP Higgs bound.
As in the other panels of this figure, the regions
currently excluded by $\bmm$ are also excluded by $b \to s \gamma$.
We note that, as $M_{in}$ decreases, the bulk and focus-point regions are
moving to larger $m_{1/2}$ more rapidly than the LEP Higgs constraint,
resulting in a larger WMAP-preferred region at small $m_{1/2}$ and
$m_0$. At the same time, however, the upper funnel wall is moving to
smaller $m_{1/2}$, causing the region between the focus point and the
upper funnel wall (where the relic density is too large) to shrink.

To illustrate how the relic density changes with $m_{1/2}$ and its
sensitivity to various interactions, we follow the evolution of the relic density for
$M_{in} = 10^{15}$~GeV at a fixed value of $m_0 = 1000$ GeV.  At very
low $m_{1/2} < 240$ GeV, the electroweak symmetry breaking conditions
would impose an unphysical solution for the weak scale value of the Higgs
mass parameter, so this region of the plane is excluded, as discussed
in Section~\ref{sec:renorm}. Near the boundary of the excluded
region, $\mu \lesssim m_{1/2}$, so the LSP has a strong higgsino
component and annihilations to light fermions keep the relic density
low. As one moves to larger $m_{1/2}$, the bino
component increases, causing the relic density to increase
accordingly, though it remains below the WMAP-preferred range. 
At $m_{1/2} = 244$~GeV, the $\chi\chi \goto W^+W^-$
threshold is reached and the relic density decreases dramatically,
only to start rising again once the threshold is passed. By
$m_{1/2} = 280$~GeV, the LSP has become bino-like, though it still has
substantial higgsino components.

Near $m_{1/2} = 325$ GeV, the relic density has risen to the range
prefered by the WMAP measurements, and continues to increase until it
exceeds the WMAP range. The thinness of the WMAP strip indicates the
rate at which the relic density is increasing, reaching its peak
value near $m_{1/2} = 500$ GeV. As $m_{1/2}$ increases further, one approaches
the broad $(H,A)$ pole region, where s-channel annihilations
cause the relic density to decrease dramatically. Thus, the upper
funnel wall appears near $m_{1/2} = 750$~GeV, and the relic density then
continues to plummet until $m_{\chi}$ becomes large enough that the
pole has been passed, at which point the relic density again
increases until it falls within the WMAP range for a third time near
$m_{1/2} = 1080$ GeV, forming the lower wall of the funnel region. As
$m_{1/2}$ increases further, the relic density of neutralinos becomes too
large to be compatible with the WMAP measurement. Near the border of
the $\stau$ LSP region, the relic density decreases due to enhanced
$\chi - \stau$ coannihilations, however the effect is not sufficient to
bring it down to the WMAP range. All values of $m_{1/2}$ to the right
of the lower funnel wall are excluded by the large relic density of neutralinos.

When $M_{in}= 10^{14.5}$ GeV, the focus-point region and upper
funnel wall merge fully to form an island of acceptable relic
density, extending from $(m_{1/2}, m_0) \sim (400, 850)$ GeV to large $m_0$, parallel to
the lower funnel wall, and with a width of $\sim 200$ GeV at its
broadest point.

\begin{figure}
\begin{center}
\mbox{\epsfig{file=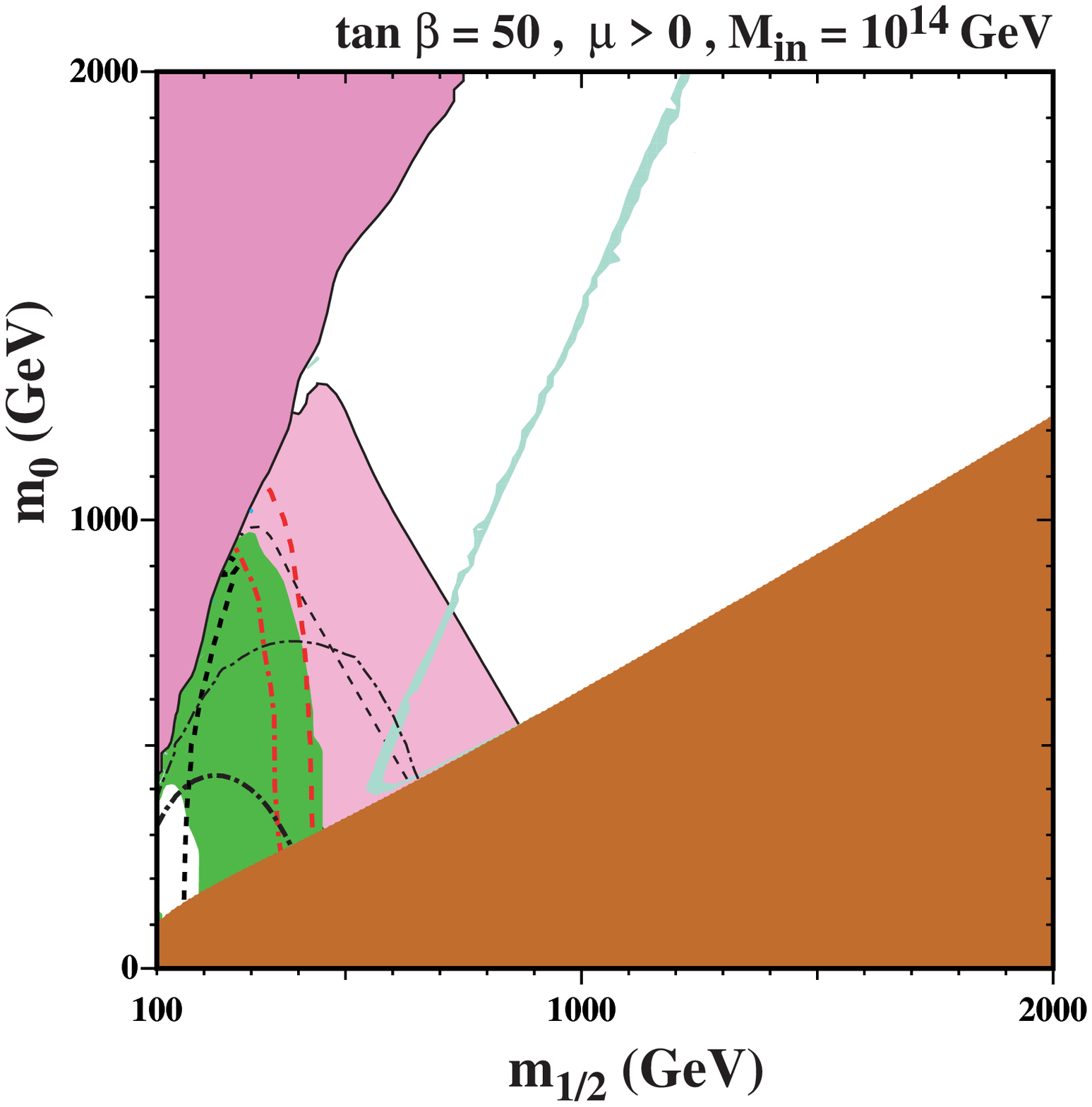,height=7cm}}
\mbox{\epsfig{file=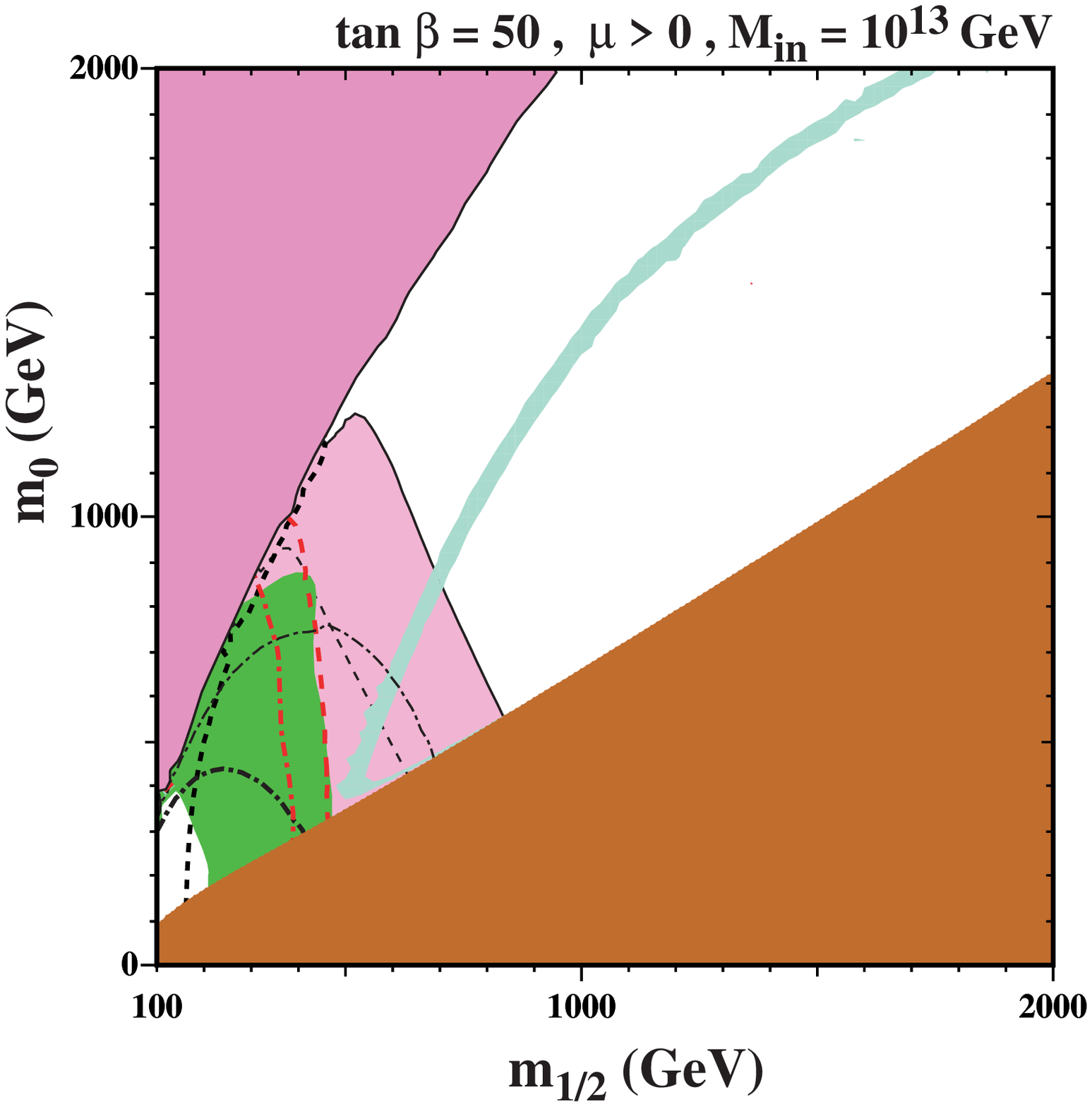,height=7cm}}
\end{center}
\begin{center}
\mbox{\epsfig{file=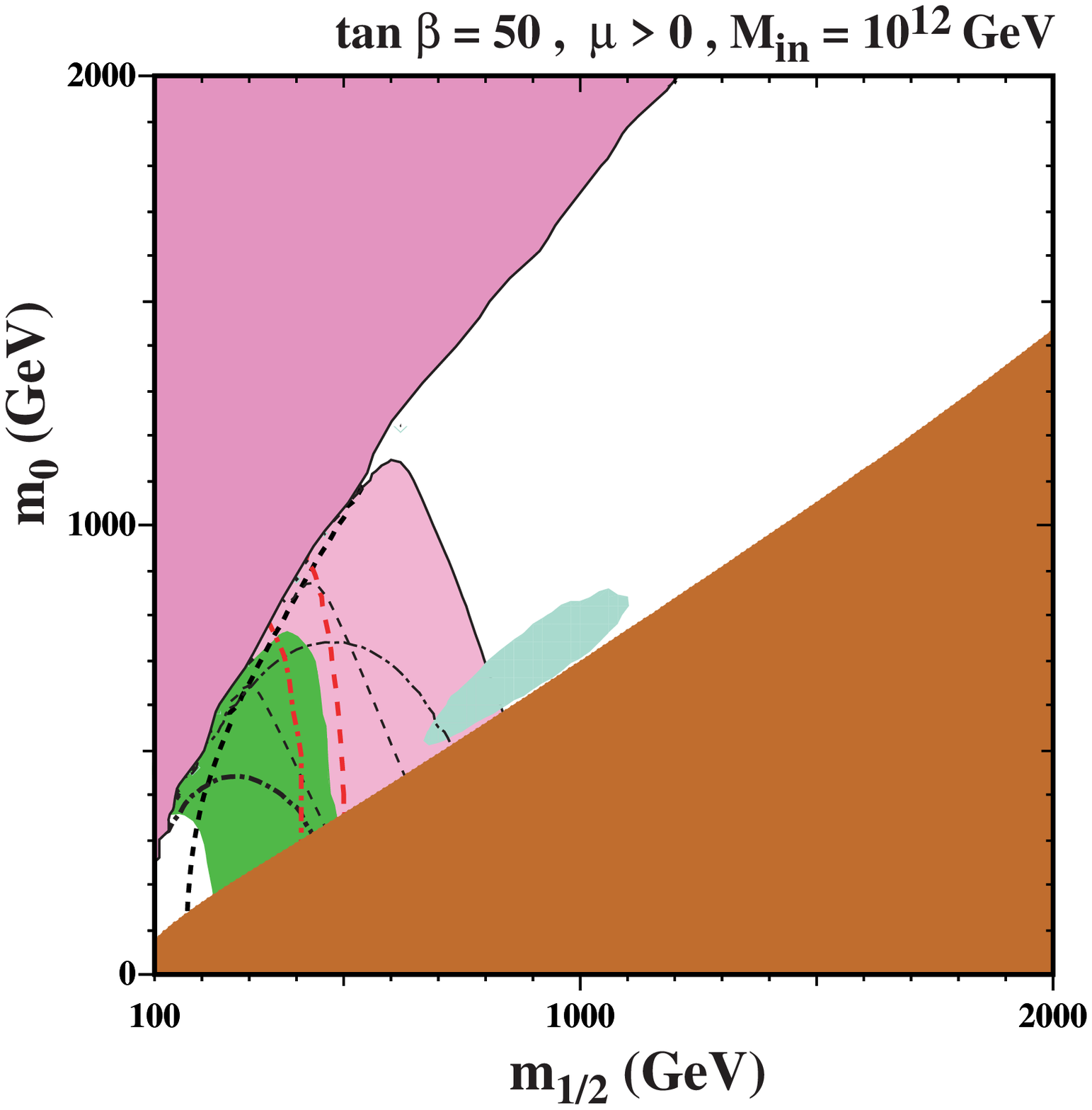,height=7cm}}
\mbox{\epsfig{file=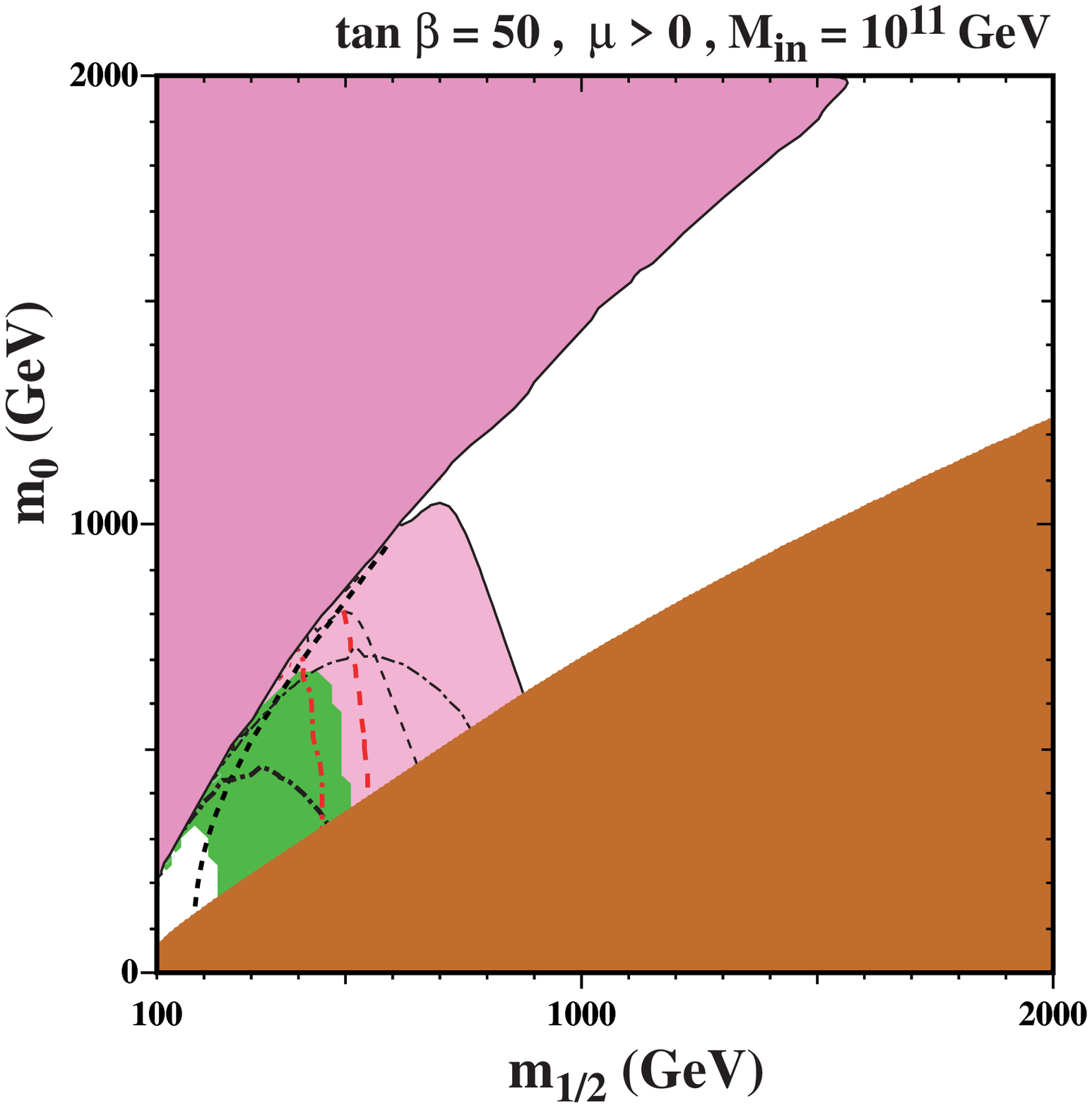,height=7cm}}
\end{center}
\caption{\it
Examples of $(m_{1/2}, m_0)$ planes with $\tan \beta = 50$ and 
$A_0 = 0$ but with  different values of $M_{in}$.
(a) $M_{in} = 10^{14}$~GeV, 
(b) $M_{in} = 10^{13}$ GeV, 
(c) $M_{in} = 10^{12}$ GeV and (d) $M_{in} = 10^{11}$ GeV.
The various contours and shadings are the same as for Fig.~4.}
\label{fig:mint502}
\end{figure}

In Fig.~\ref{fig:mint502}, as in the $\tbt = 10$ scenario, we see
the electroweak vacuum condition creep further down into the plane,
as $M_{in}$ is further reduced.
The $\stau$ LSP region also retreats to smaller $m_0$,
because of the $M_{in}$ dependences of the sparticle masses discussed in 
Section~\ref{sec:renorm}. When the universality scale is $M_{in} = 10^{14}$~GeV, 
as seen in panel (a) of Fig.~\ref{fig:mint502}, this island has submerged
and disappeared as enhanced annihilations to $b\bar{b}$ and $\tau \bar{\tau}$
dominate even for $2m_{\chi_1} < m_A$. Coannihilations of $\chi_0$ with
$\chi_i$, where $\chi_i$ denote the second- and third-lightest neutralinos,
also play a significant role in the smallness of the relic density in
this region. The only values of $m_{1/2}$ and $m_0$ for which the relic density of neutralinos is
in agreement with the WMAP measurement are in the thin strip that had been the
lower funnel wall, and a narrow coannihilation strip adjacent to the $\stau$ LSP region.  
To the left of the residual funnel strip, the relic density is
below the WMAP value, whereas this value is exceeded in the `vee' between the funnel 
and coannihilation strips at large $m_{1/2}$.  At $M_{in} = 10^{14}$ GeV, all values of
$m_{1/2} > 1230$ GeV are excluded for $m_0 < 2000$ GeV. 

In panel (b) for $M_{in} = 10^{13}$~GeV, what is left of the lower funnel wall is also beginning to
curve down.  This is the same general behavior we observed in the
$\tbt = 10$ case.  As the universality scale is slightly reduced, to $M_{in}
= 10^{12.5}$~GeV (not shown), this strip bends down into the $\stau$ LSP
region at $(m_{1/2}, m_0) \sim (2000, 1450)$~GeV. For $M_{in} = 10^{12}$ GeV, 
as seen in panel (c) of Fig.~\ref{fig:mint502}, 
there remains only a small ellipse where the relic density falls in the region
preferred by WMAP. The rest of the plane not excluded by the
electroweak vacuum condition or the charged LSP constraint has a relic density
of neutralinos smaller than that required by WMAP~\footnote{We stress again
that such regions are not excluded, provided there is another source
of cold dark matter in the Universe.}.  This last remaining WMAP island
evaporates as the universality scale is decreased to $10^{11}$~GeV, 
as seen in panel (d) of Fig.~\ref{fig:mint502}, at
which point the entire plane is disfavoured, in the sense that some additional
source of cold dark matter would be required.

\subsection{Non-Zero $A_0$
\label{sec:A0}}

To this point, we have considered all trilinear soft
supersymmetry-breaking parameters to be zero
at the unification scale, $A_0 = 0$. Here we limit
ourselves to a brief discussion of $A_0 \ne 0$ as preparation for the discussion
of mirage-mediation models in the next Section.

If $A_0 > 0$, the RGEs generate
correspondingly larger trilinear couplings at the weak scale. In
addition, since the large loop corrections to $\mu$ depend on the values of
the trilinear couplings, there is also an increase in $\mu$. We
therefore expect, based on the discussion in Section \ref{sec:renorm},
that the region excluded by the electroweak vacuum condition decreases
with increasing $A_0$. Other striking differences in the $(m_{1/2},m_0)$ plane are in the
constraints on the Higgs mass and the $b \goto s \gamma$ rate. While
the LEP Higgs constraint is dramatically relaxed for larger $A_0$, the
region excluded by $b \goto s \gamma$ increases in size, becoming the
dominant constraint for low $m_{1/2}$. Furthermore, since the
off-diagonal elements of the squark mass matrix contain terms
proportional to the negative of the trilinear couplings,
when $A_0$ is large these off-diagonal contributions can become large
enough to drive the lightest stop quark mass below the LEP bound.  
As a result, we see a new excluded region emerge at low $m_{1/2}$ and $m_0$, where the lighter
stop has $m_{\stop} < 220$ GeV \cite{stop}. 

For $A_0$ negative, the changes to the constraints discussed above are
quite predictable.  In this case, the RGE's generate correspondingly smaller
weak scale trilinear couplings, resulting in a universally smaller
$\mu$.  The LSP is then more more Higgsino-like over the whole
plane. The LEP Higgs bound is strengthened, and the $b \goto s
\gamma$ rate becomes an insignificant constraint.

The regions of the plane where the relic density of neutralinos is in
the measured range also change shape for $A_0 \neq 0$. In general,
these changes can be ascribed to one of two effects. First, in
addition to the $\stau\chi$ coannihilation strip, there may be an
additional $\stop\chi$ coannihilation strip, where the lighter stop is
degenerate with the neutralino LSP.  This feature is common in
scenarios with large $A_0$ and both the $\stop\chi$ coannihilation
strip and the excluded light stop region move further into the plane
as $A_0$ is increased.  Secondly, we recall that the composition
of the LSP depends on the ratio of $\mu$ to $M_1$, the LSP being
bino-like when $M_1$ is small compared to $\mu$ and Higgsino-like if $\mu$ is small compared
to $M_1$, as shown in panel (d) of Figure \ref{fig:masses}. Since $\mu$ is
enhanced everywhere in the plane when $A_0 > 0$, we expect the LSP to
be generically more bino-like than when $A_0 = 0$. Similarly, we
expect the LSP to be generically more Higgsino-like when $A_0 < 0$. For $M_{in} \approx
M_{GUT}$ (not pictured), the LSP is strongly bino-like over most of the plane, so the main
effects of $A_0 \neq 0$ are the above-mentioned modifications in the constraints, 
and the appearance of the $\stop\chi$ coannihilation strip for large
positive $A_0$.
 
For lower unification scales, however, the LSP has
more substantial Higgsino components, becoming Higgsino-dominated over much
of the plane for very low $M_{in}$. Larger $\mu$ means that the LSP will
remain bino-like even for larger values of $M_1$, so in scenarios
with $A_0 >0$ the LSP is more bino-like and the heavier neutralinos
with large Higgsino components are even heavier than when $A_0 =
0$. These differences are clear at low $M_{in}$, when the LSP is becoming Higgsino-like over much
of the plane when $A_0 = 0$ but is still bino-like when $A_0$ has a
sufficiently large positive value.  In panel (a) of Fig.~\ref{fig:mirage}, we
show the $(m_{1/2},m_0)$ plane for $\tbt = 10$, $M_{in}=10^{12}$ GeV,
and $A_0 = 1000$ GeV. We note the similarity to panel (d) of 
Fig.~\ref{fig:mint}, where $M_{in}= 10^{12.5}$. When $A_0>0$, smaller
values of $\mu$ appear only at values of $M_{in}$ that are lower than in
the $A_0 = 0$ cases previously discussed. In the same way, the $A_0 < 0$ case
tends to mimic the effect of larger $M_{in}$.  With respect to
the relic density of neutralinos, there is some degeneracy in the
parameters $M_{in}$ and $A_0$ for regions of the $(m_{1/2},m_0)$ plane
far from the $\stop\chi$ coannihilation strip.

We note that $A_0 \propto M$, where $M = m_{1/2}$ or $m_0$, is also a viable
possibility, the consequences of which, in light of the above
discussion, are easily understood.  In these cases, for small $M$, the
plane will be similar to the $A_0 = 0$ case, while at larger $M$, the
changes described above will be increasingly evident. A complete
discussion of $A_0 \ne 0$ or non-universal $A_0$ is beyond the scope
of this study.


\subsection{Mirage-Mediation Models}
\label{sec:mirage}

Models in which supersymmetry breaking occurs through some combination
of modulus and anomaly mediation are among those
characterized by the apparent unification of gaugino and scalar mass parameters at an
intermediate scale.  As a result, these models have been termed mirage-mediation 
models~\cite{mixed}~\footnote{Such models are motivated, e.g., by the KKLT framework~\cite{KKLT}.}, 
and the unification scale, the mirage messenger
scale, is estimated to be $\sim 10^{10}-10^{12}$~GeV. 
One distinctive feature of these scenarios is that the gaugino and scalar masses run both above
and below the unification scale.  Here, we discuss briefly
the effect on our results of the additional running of the masses above the unification scale.

The use of the RGEs to run the masses down from the input scale to the weak scale 
is unchanged, and the procedure for calculating the weak-scale observables is
unchanged, regardless whether the soft supersymmetry-breaking mass
parameters run above the unification scale. The chief
difference derives from the fact that the value of $\mu$ is fixed by
the electroweak vacuum conditions, which include a large dependence on
the trilinear couplings as discussed in the previous section. 
When the trilinear couplings run from
the GUT scale, becoming larger as the energy scale decreases, they
attain larger weak-scale values than would be possible with running
only below $M_{in}$. Therefore, in mirage-mediation models $\mu$ receives a
large contribution from the exceptionally large values of the trilinear
couplings at the weak scale. The resulting picture for mirage-mediation
models is similar to what one would expect from the GUT-less
cases with $A_0 \neq 0$, as discussed above. It should be noted that
the trilinear couplings in mirage-mediation scenarios, as well as the
other soft SUSY-breaking parameters, are specified at the GUT scale
based on the particular mixture of modulus and anomaly
mediation. The soft SUSY-breaking parameters are taken to be
proportional to each other, with constants of proportionality
determined by the modular weights and other considerations \cite{mixed}. For simplicity, we
consider only $A_0 = 0$ at the GUT scale.

\begin{figure}
\begin{center}
\mbox{\epsfig{file=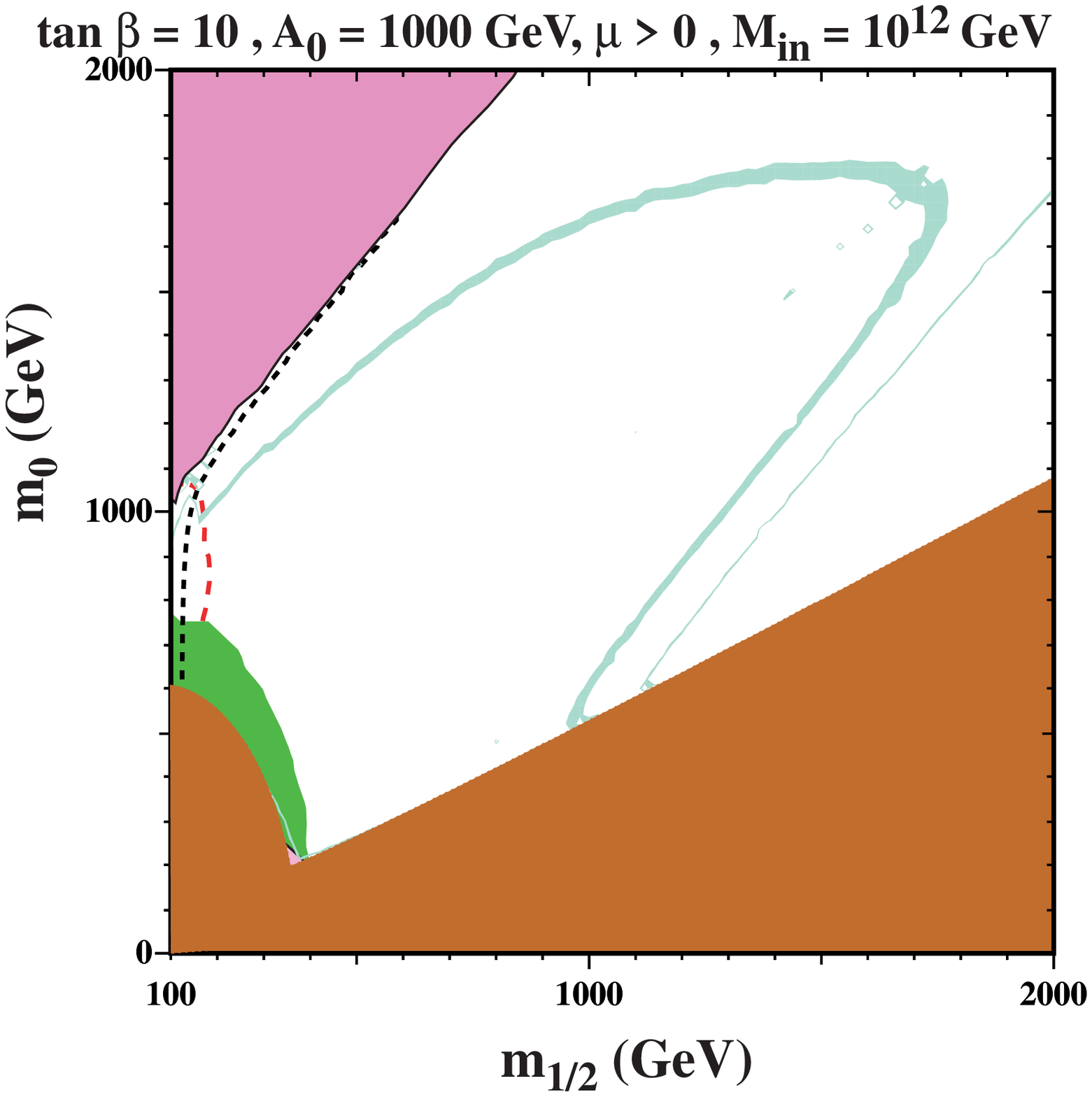,height=7cm}}
\mbox{\epsfig{file=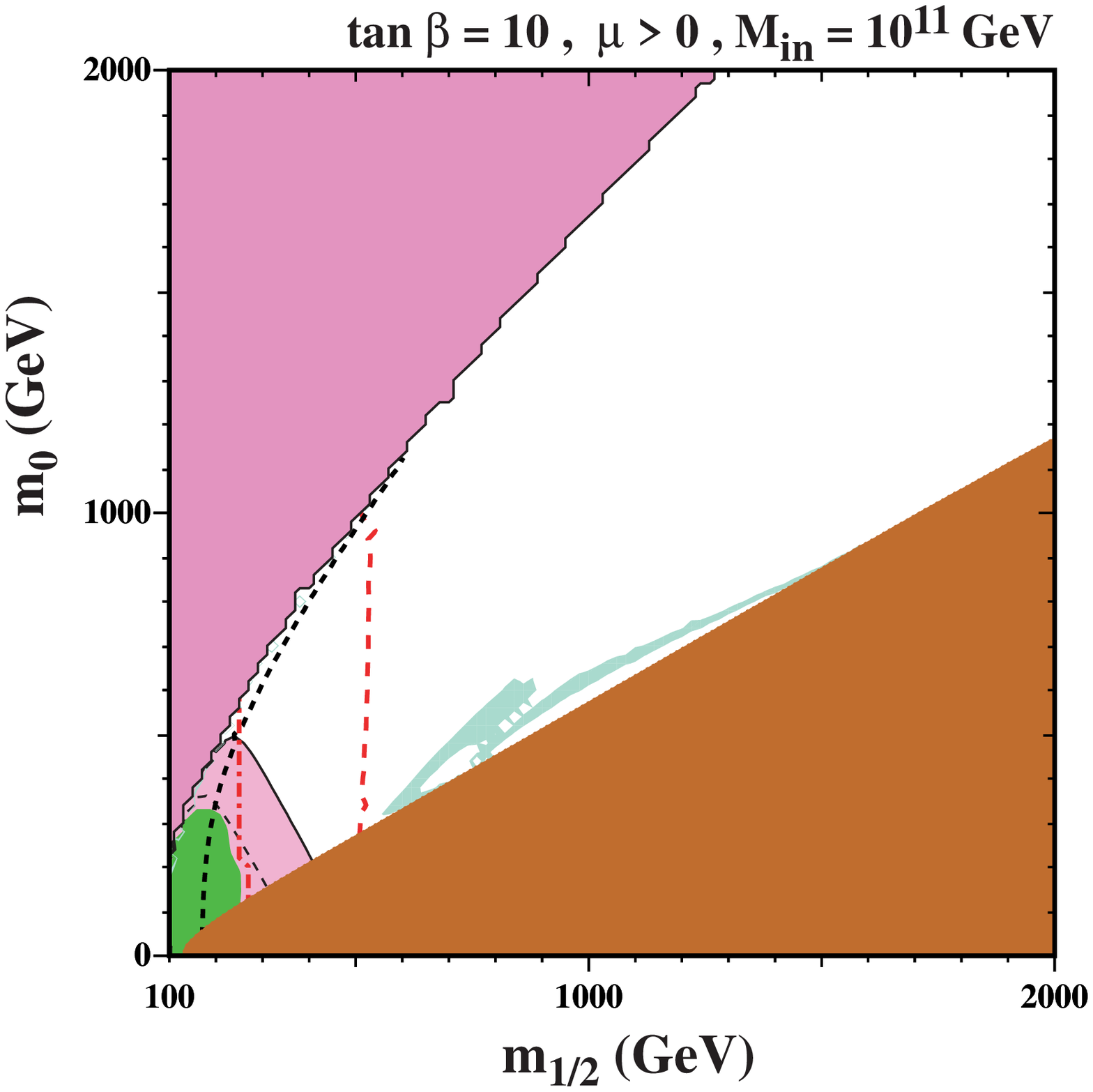,height=7cm}}
\end{center}
\begin{center}
\mbox{\epsfig{file=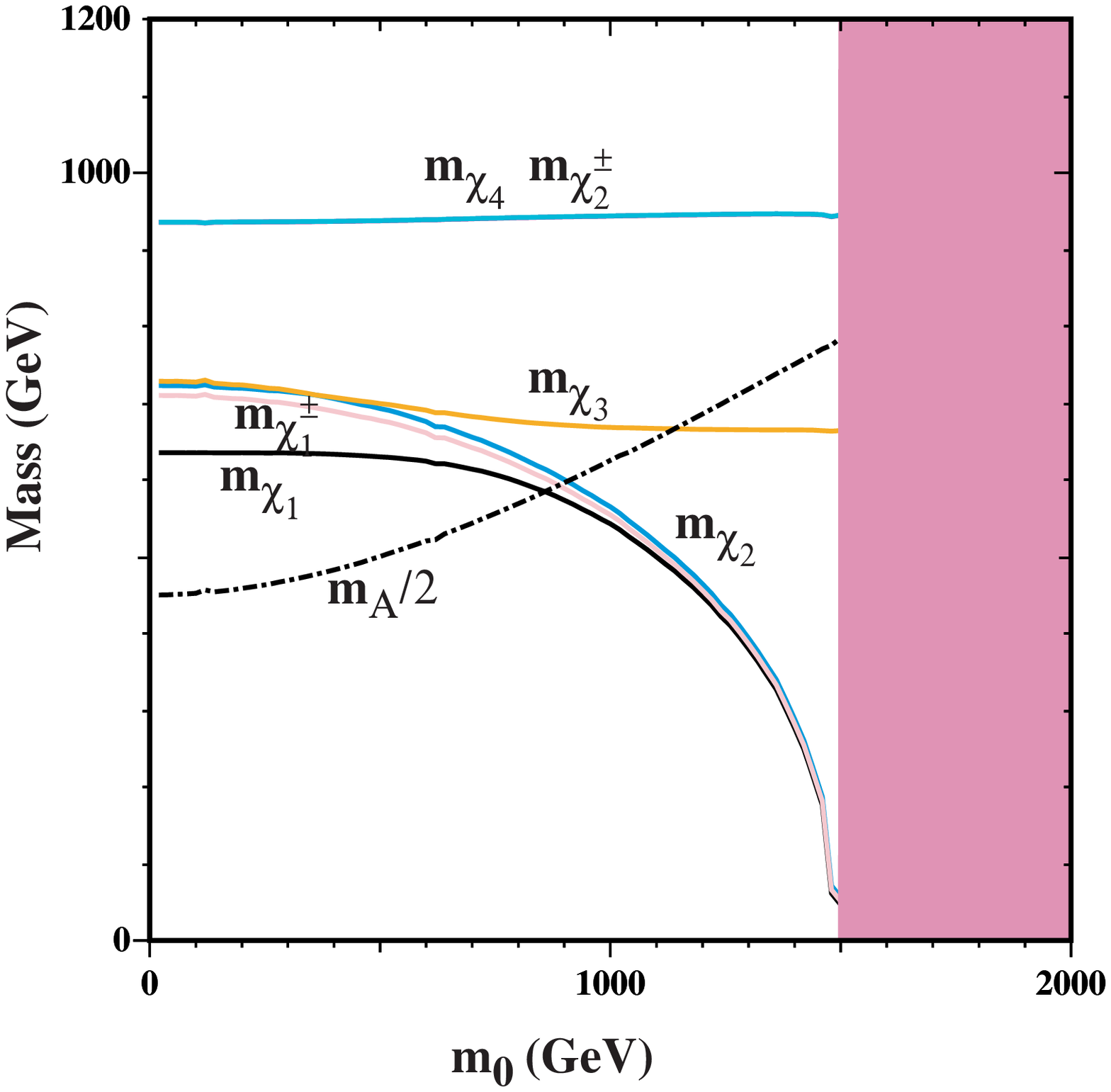,height=7cm}}
\mbox{\epsfig{file=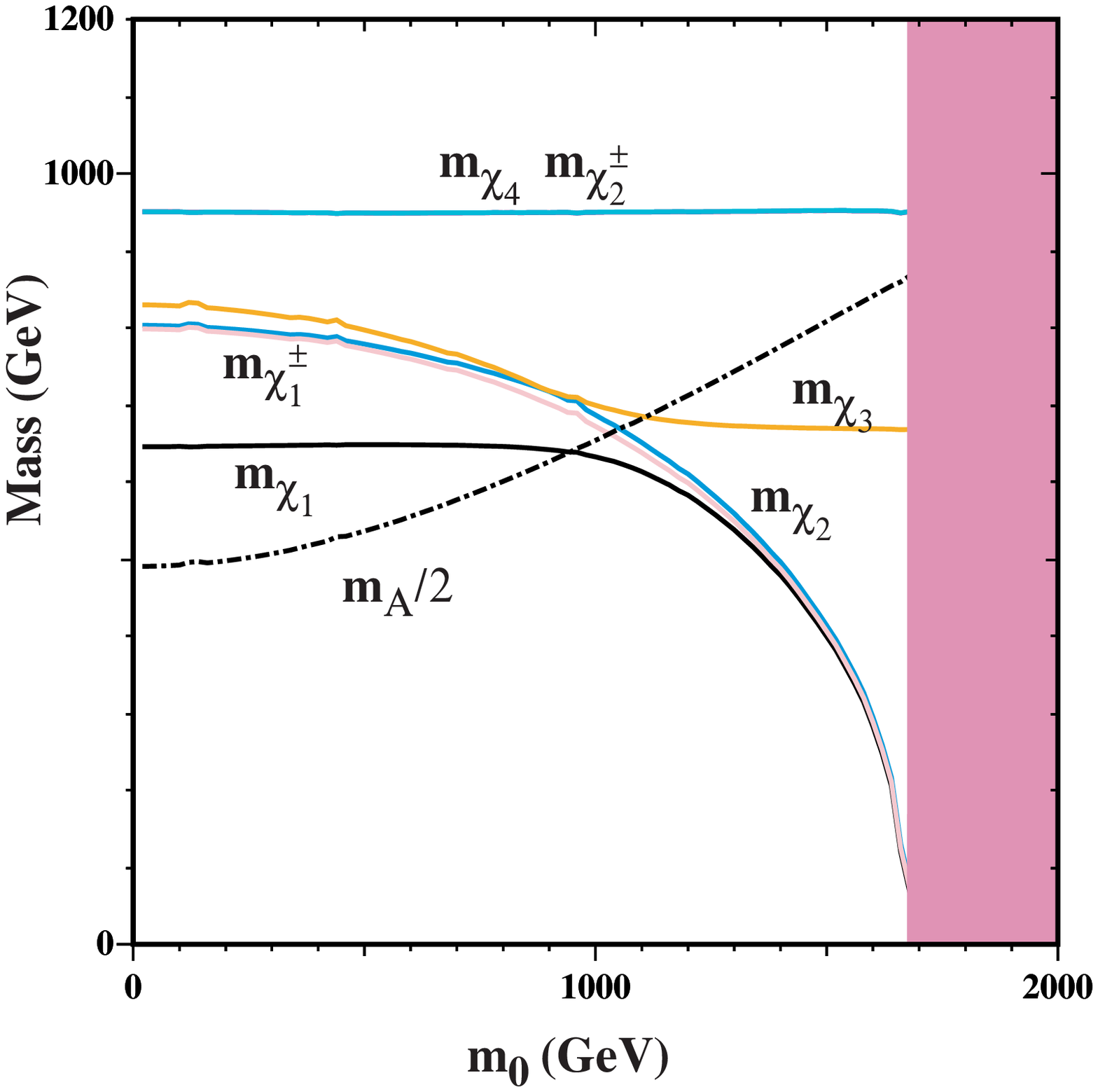,height=7cm}}
\end{center}
\caption{\it Panel (a) shows the $(m_{1/2},m_0)$ plane for the GUT-less
case with $\tbt=10$, $M_{in}=10^{12}$ GeV, and $A_0=1000$ GeV. Panel (b) displays a scenario
similar to that found in mirage-mediation
models, where the soft supersymmetry-breaking parameters are universal at
$M_{in} = 10^{11}$ GeV, but run both above and below this scale. The weak-scale
values of the neutralino and chargino masses, as well as the
pseudoscalar Higgs mass and $\mu$, are shown in panel (c) for the usual
GUT-less case with $A_0=0$ and $M_{in}=10^{11}$ as shown in
panel (b) of Fig.~\ref{fig:mint2}. Panel (d) shows the same
information as panel (c) for the mirage-mediation case.}
\label{fig:mirage}
\end{figure}

In panel (b) of Fig.~\ref{fig:mirage} we
show the $(m_{1/2},m_0)$ plane with running of the
gaugino and scalar masses both above and below the unification scale for $M_{in} =
10^{11}$ GeV and $\tbt=10$. There is a broad region of acceptable relic density lying just above the
excluded $\stau$ LSP region. For comparison, in the standard GUT-less
case for $M_{in}=10^{11}$ GeV shown in panel (b) of Fig.~\ref{fig:mint2}, 
as discussed already in Section~\ref{sec:lowtanb}, the relic density of neutralinos is below the 
WMAP 2-$\sigma$ range throughout the plane, 
except in the small island just barely in view at $m_{1/2} = 2000$ GeV. 

There are a few important differences worthy of note.
First, the value of $\mu$ all over the plane is universally larger in the
mirage-mediation scenario than in the cases discussed previously in this
paper, which is attributed to the running of $A_0$ from the GUT
scale rather than $M_{in}$. As a result, we expect the boundary of the region excluded
by the electroweak vacuum conditions to be pushed back up into the
upper left corner of the plane, as is seen. A second important consequence concerns the 
composition of the
LSP. Recalling that the LSP is bino-like as
long as $M_1$ is much smaller than $\mu$. The fact that $\mu$ is larger in the
mirage-mediation case
implies that the cross-over when $M_1 \approx \mu$ takes place at a lower unification
scale than was found in panel (d) of Fig.~\ref{fig:masses}. In fact, the
LSP is bino-like over most of the plane in the mirage-mediation case
shown in panel (b) of Fig.~\ref{fig:mirage}, whereas it has
large Higgsino components for much of the standard GUT-less plane for
the same value of $M_{in}$. Similarly, the heavier neutralinos, which have large Higgsino
components, are even heavier due to the enhancement in $\mu$ in
mirage-mediation models. This effect can be seen clearly by comparing panels (c) and (d)
of Fig.~\ref{fig:mirage}.



\section{Neutralino-Nucleon Cross Sections}

Direct searches for dark matter particles such as the Cryogenic Dark
Matter Search (CDMS) \cite{cdms} and other
experiments look for evidence of weakly-interacting massive
particles (WIMPS) through scattering on nuclei.  In this section, we
present the predictions for neutralino-nucleon scattering cross
sections in the scenarios discussed above  \cite{etal} - \cite{bot2}. 

The low-energy effective
interaction Lagrangian for elastic $\chi$-nucleon scattering can be written as 
\beq
L = \alpha_{2i} \bar{\chi} \gamma^{\mu}\gamma^5 \chi \bar{q}_i
\gamma_{\mu}\gamma^5 q_i + \alpha_{3i} \bar{\chi}\chi\bar{q}_i q_i \; ,
\eeq 
where terms that make velocity-dependent contributions to the cross
section have been neglected, and the constants $\alpha_{2i}$ and
$\alpha_{3i}$ are defined as in Ref.~\cite{EFlO1}. In computing the scalar cross section,
we have assumed the pi-nucleon $\Sigma$ term to be 64 MeV (see \cite{eoss8} for the
sensitivity of the elastic cross section to this assumption).
Summation over the quark generations is
implied, with up- and down-type quarks labeled by the subscript
$i$. The cross section can be broken into a spin-dependent part
arising from the term proportional to $\alpha_{2i}$ and a
spin-independent (scalar) part from the term
proportional to $\alpha_{3i}$. The spin-dependent cross section is, in
general, larger than the scalar cross section. However, since the whole
nucleus participates coherently in spin-independent interactions, it is primarily
the scalar cross section that is probed by current direct-detection experiments.
On the other hand, the spin-dependent scattering cross section on the proton plays
an important role in the capture and annihilation rates inside the Sun.

\begin{figure}
\begin{center}
\mbox{\epsfig{file=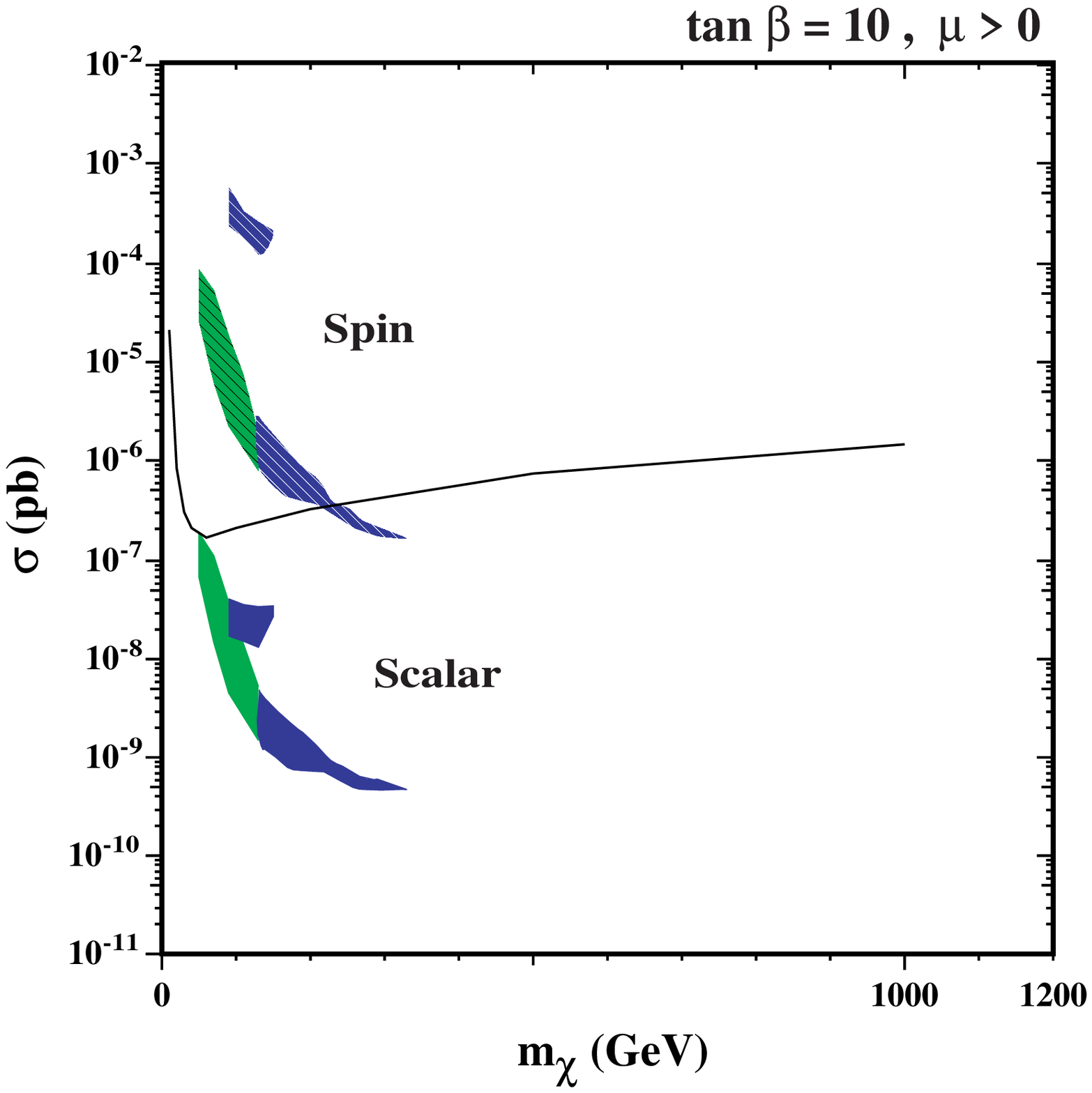,height=7cm}}
\mbox{\epsfig{file=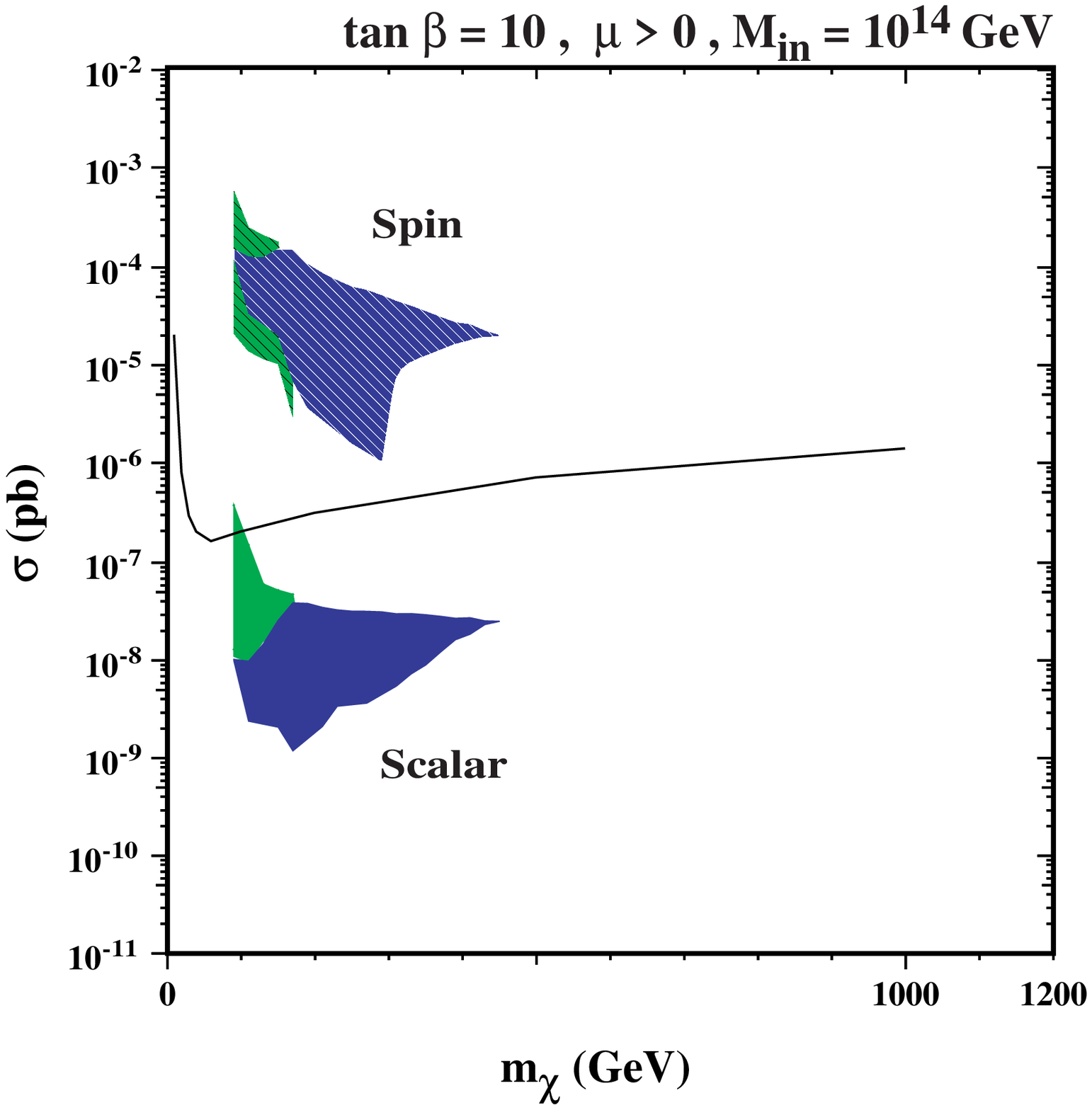,height=7cm}}
\end{center}
\begin{center}
\mbox{\epsfig{file=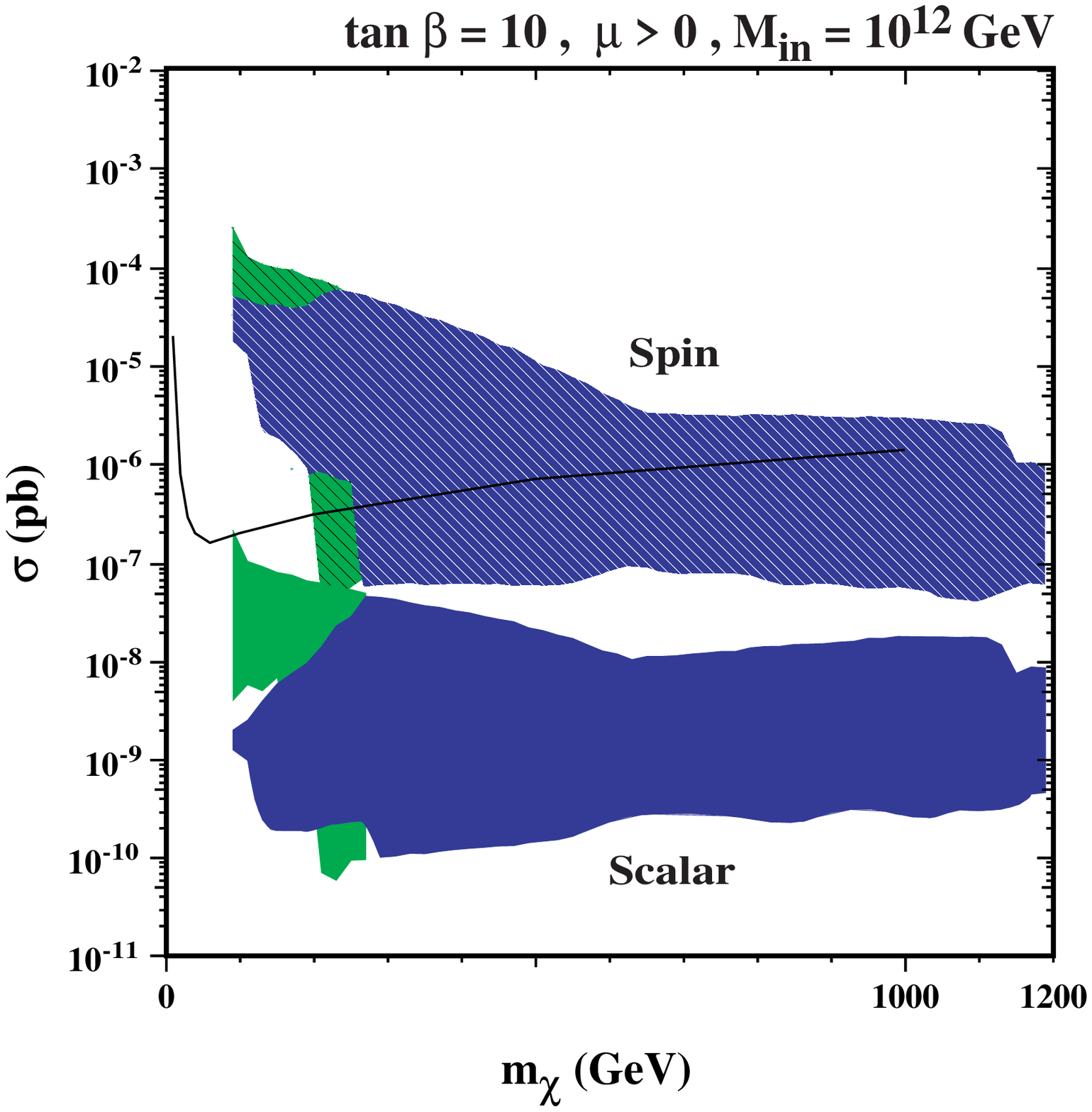,height=7cm}}
\mbox{\epsfig{file=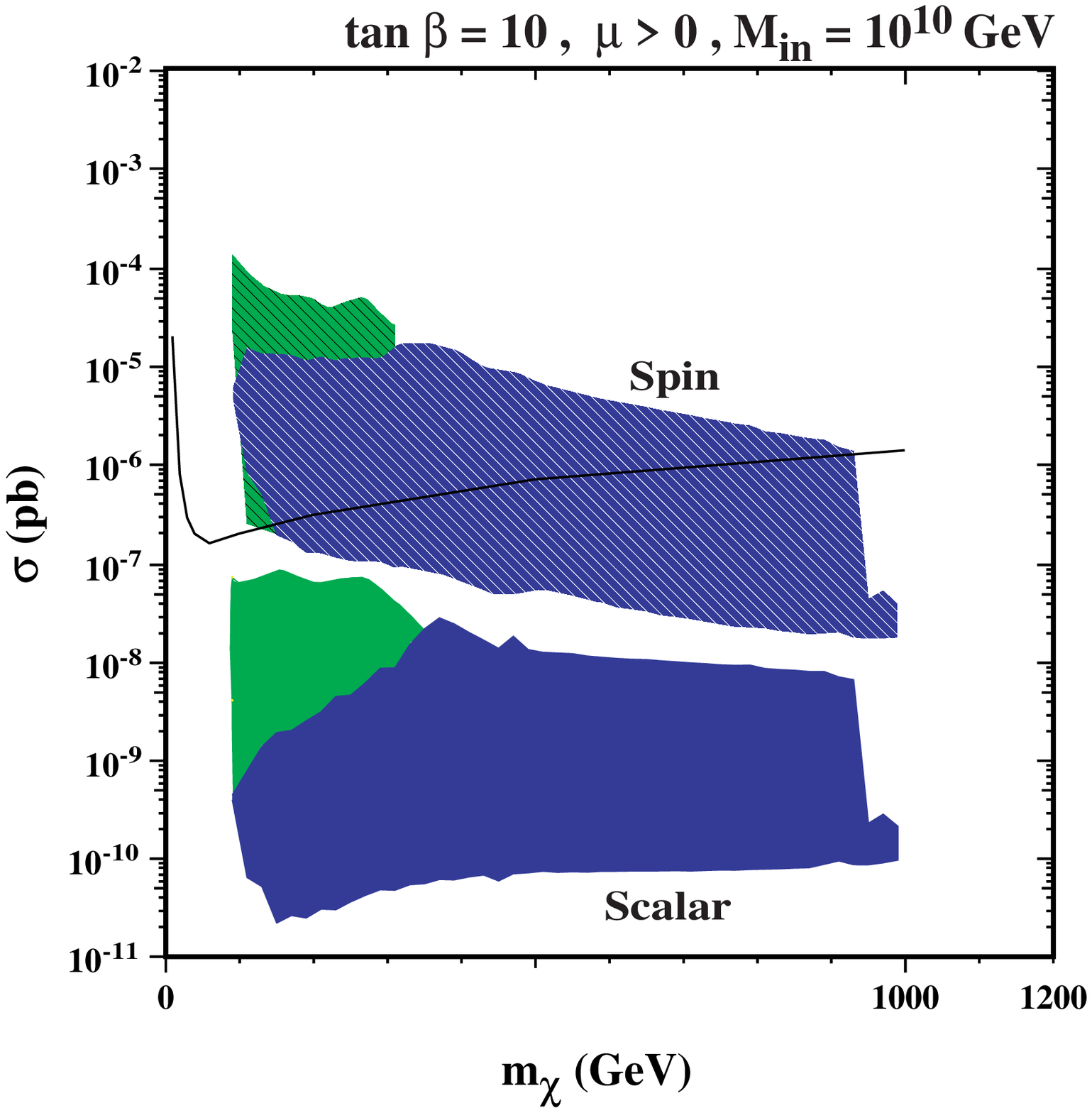,height=7cm}}
\end{center}
\caption{\it Neutralino-nucleon cross sections as functions of the
neutralino mass for $\tan \beta = 10$ and 
$A_0 = 0$ but with  different values of $M_{in}$.
(a) $M_{in} = M_{GUT} \approx 2 \times 10^{16}$ GeV, 
(b) $M_{in} = 10^{14}$ GeV, 
(c) $M_{in} = 10^{12}$ GeV and (d) $M_{in} = 10^{10}$ GeV.}
\label{fig:cs10}
\end{figure}

\begin{figure}
\begin{center}
\mbox{\epsfig{file=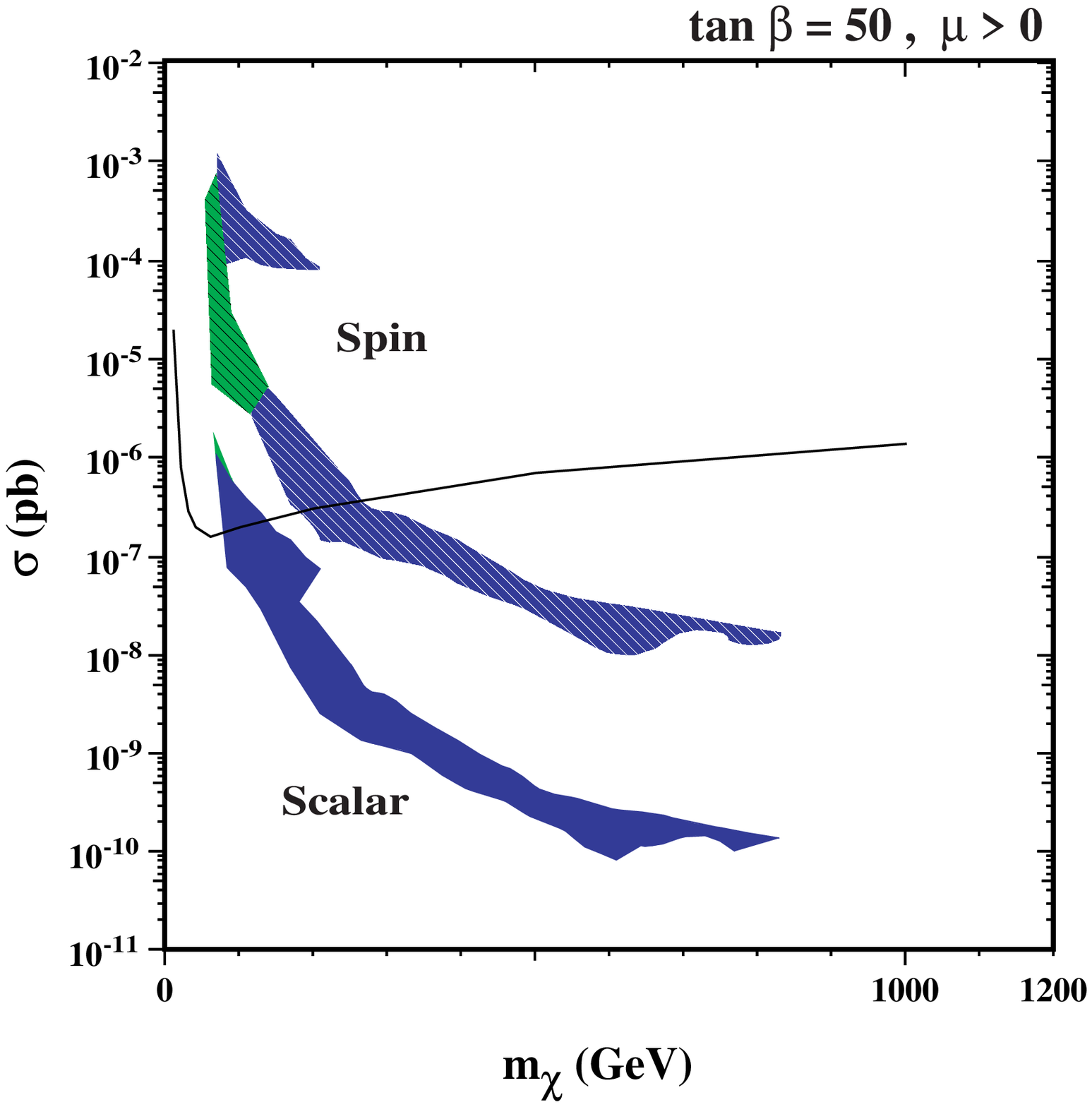,height=7cm}}
\mbox{\epsfig{file=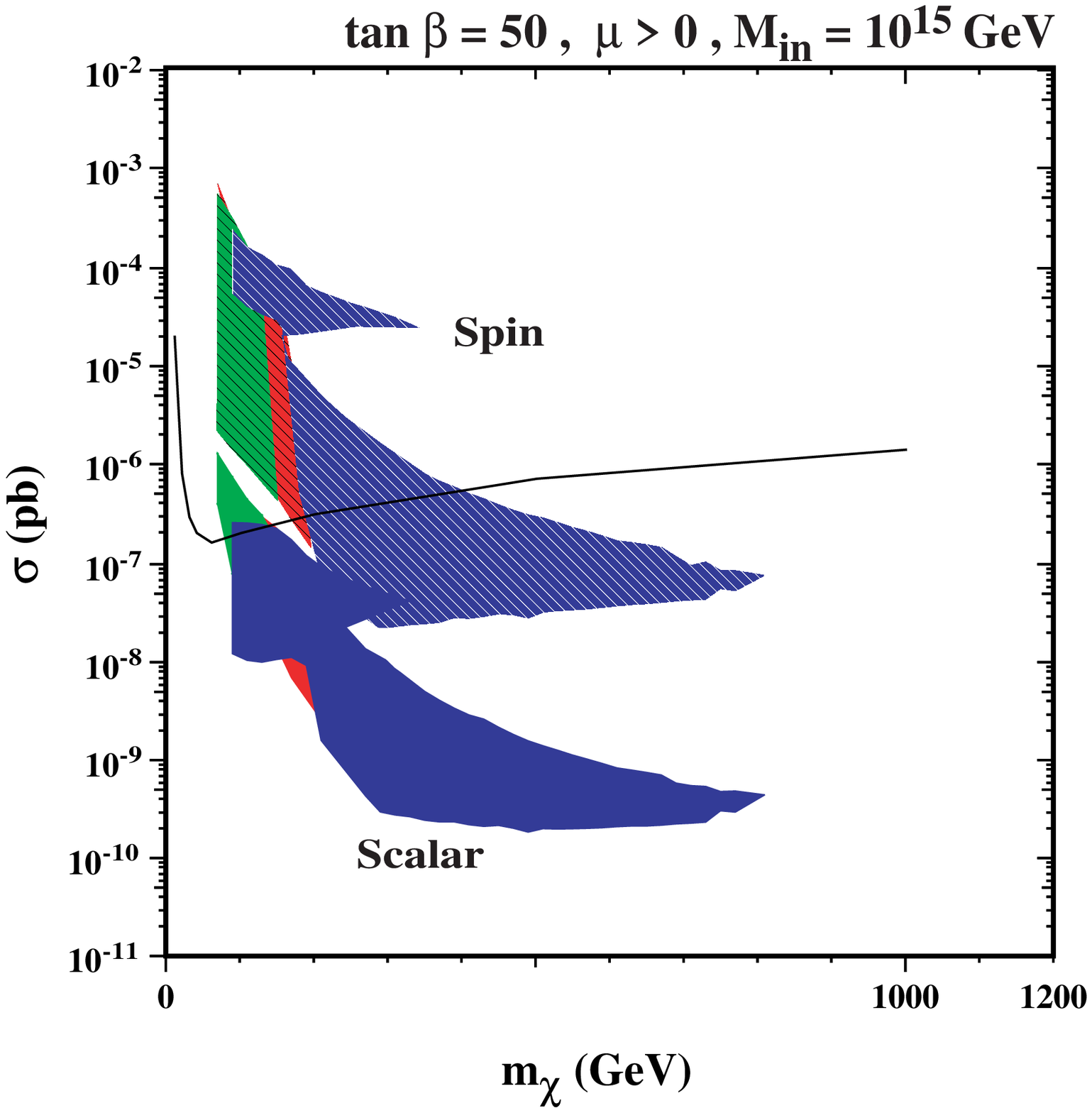,height=7cm}}
\end{center}
\begin{center}
\mbox{\epsfig{file=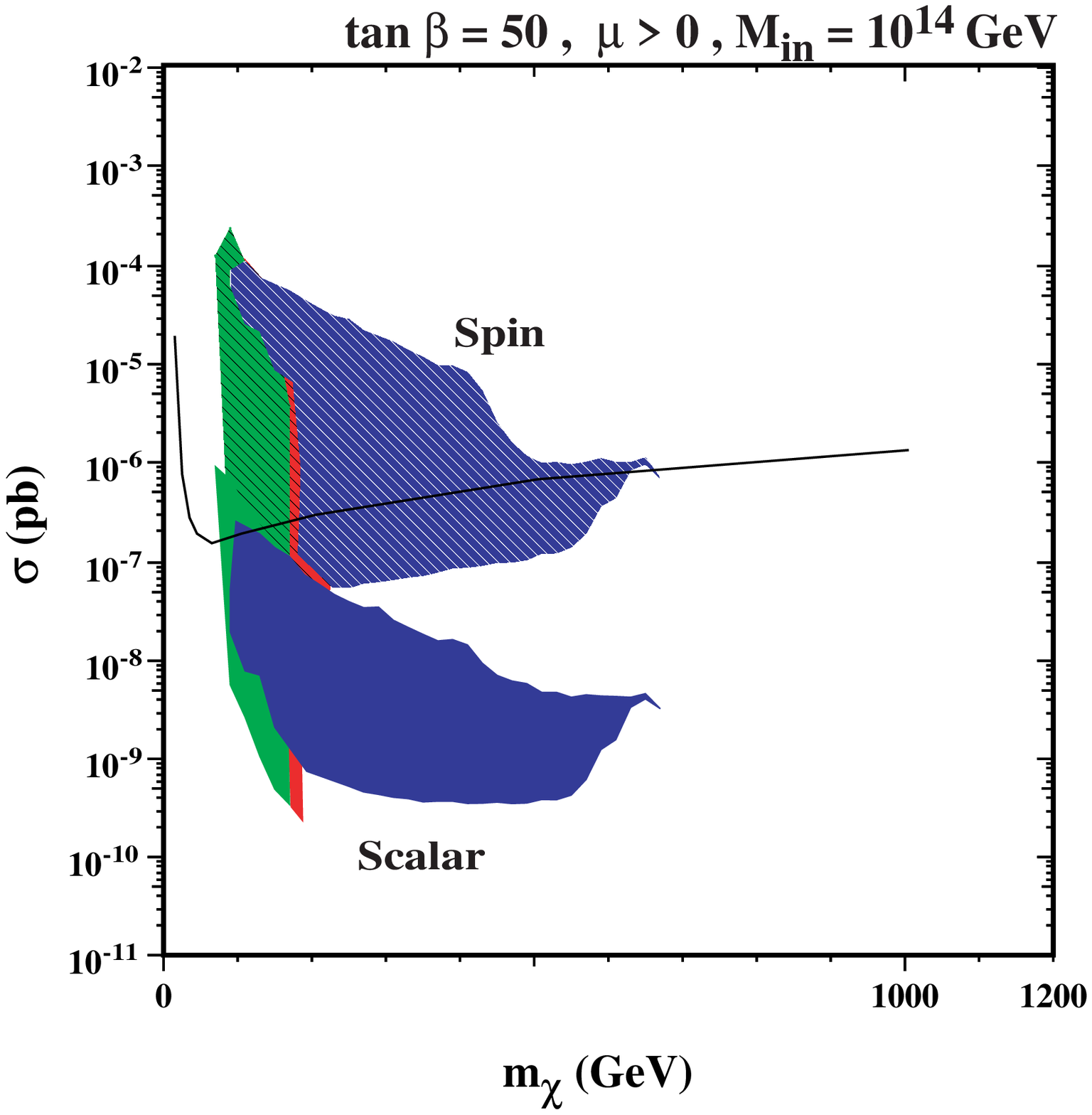,height=7cm}}
\mbox{\epsfig{file=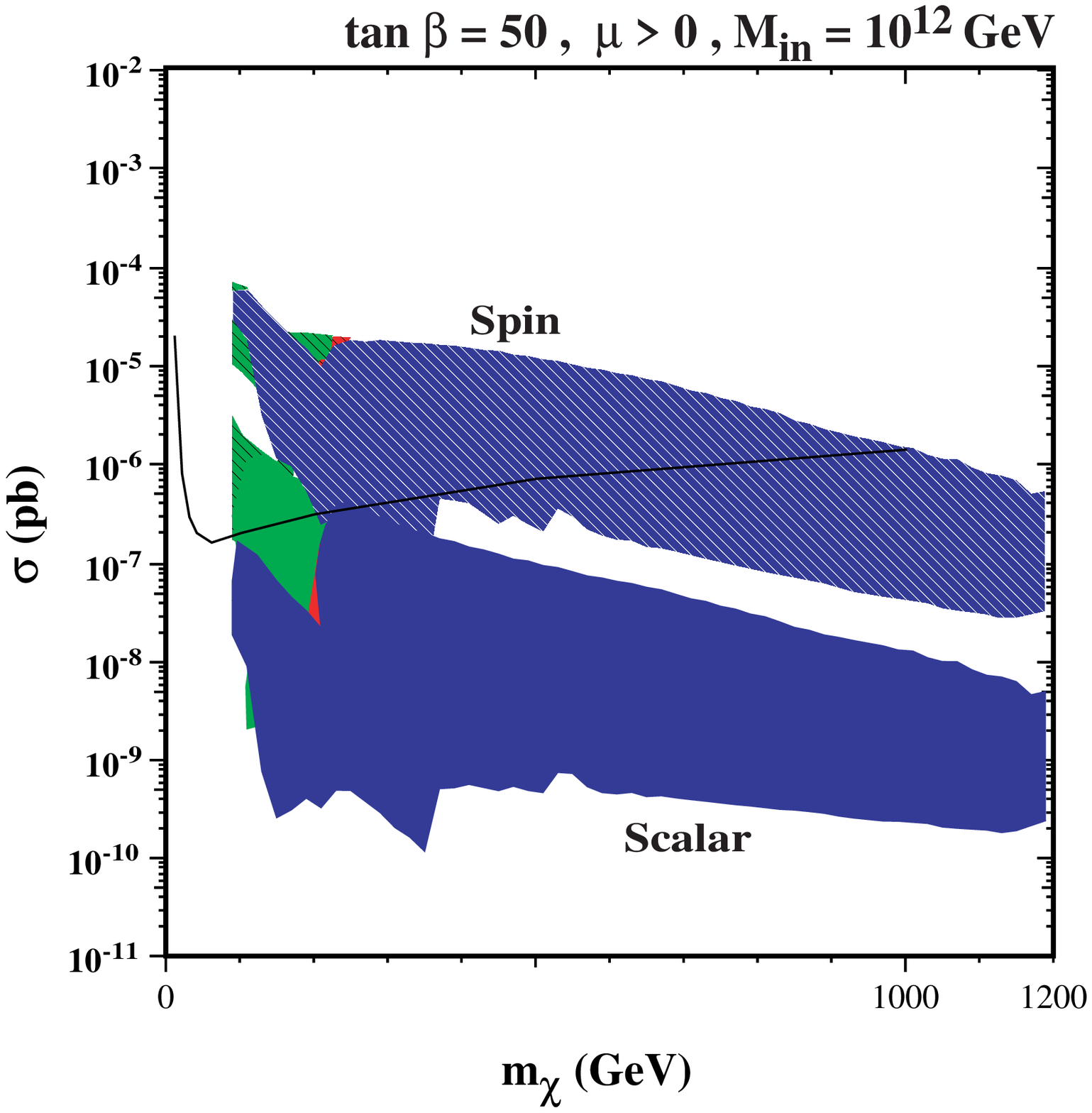,height=7cm}}
\end{center}
\caption{\it Neutralino-nucleon cross sections as a function of
neutralino mass for $\tan \beta = 50$ and 
$A_0 = 0$ but with  different values of $M_{in}$.
(a) $M_{in} = M_{GUT} \approx 2 \times 10^{16}$ GeV, 
(b) $M_{in} = 10^{15}$ GeV, 
(c) $M_{in} = 10^{14}$ GeV and (d) $M_{in} = 10^{12}$ GeV.}
\label{fig:cs50}
\end{figure}

Figs.~\ref{fig:cs10} and \ref{fig:cs50} show scatter plots of the spin-dependent 
and scalar cross sections for elastic $\chi$-nucleon
scattering.  We plot the cross
sections as functions of the neutralino mass for points in the
$(m_{1/2},m_0)$ plane where the relic density of neutralinos is less
than the 2-$\sigma$ upper limit from WMAP (as first examined in
Ref.~\cite{ckm01}) with the assumption of
universality at the GUT scale relaxed. For the cases where the
relic density is smaller than the central WMAP value, indicating that
there must be another source of astrophysical cold dark matter, we
plot the cross section scaled by the ratio of the relic density of neutralinos to the 
central density of cold dark matter inferred from WMAP measurements of the CMB. 
These results can be compared with
the direct-detection limits available from CDMS and other experiments.
In each figure, we also show the CDMS II limit for the scalar part of the
neutralino-nucleon cross section \cite{cdms2}. Current limits on the spin-dependent
cross section are $\sigma_{\chi n} \lesssim 10^{-1}$ pb~\cite{kims}, which lies
outside the range we have plotted in Figures \ref{fig:cs10} and
\ref{fig:cs50}. We require that the lightest neutralino be
the LSP and that electroweak symmetry be broken, as usual. 
The LEP constraint on the chargino mass has been applied, as
discussed in Section \ref{sec:LEPconstraints}. Different colors in
Figures \ref{fig:cs10} and \ref{fig:cs50} indicate whether the point lies within the region 
excluded by $b \goto s \gamma$ decay or the LEP Higgs mass constraint. 
The dark
blue (striped) regions are the spin-dependent (scalar) cross
sections that pass all these constraints. Lighter (green) regions in
each panel fail
the relaxed LEP Higgs constraint. At large $\tbt$, when the constraint
on the rate of $b \goto s \gamma$ becomes dominant, we show in red the regions
that fail this constraint but pass all others. 
 
The results for $\tbt = 10$ are shown in Figure \ref{fig:cs10} for
four different values of $M_{in}$. The spin-dependent cross sections
lie above the scalar cross sections in each of the four panels. As
$M_{in}$ is lowered, the number of points increases dramatically and
they spread to larger $m_{\chi}$. This
is due to the fact that the relic density over all of the
$(m_{1/2},m_0)$ plane decreases as $M_{in}$ decreases, so that less
and less of the plane is excluded by having an excess relic density~\footnote{In fact, for
$M_{in} = 10^{10}$~GeV as shown in Panel (d), the constraint on the
relic density does not exclude any points, but serves only as a scale
factor for the cross sections.}. 

We turn our attention first to the usual GUT-scale CMSSM, in which the
relic density is too large over most of the $(m_{1/2},m_0)$
plane.  Within the allowed regions for each of the spin-dependent and
scalar cross sections, we can identify two separate behaviours. First,
there is a region stretching out to $m_{\chi} \sim 350$ GeV where the cross section
may vary over as much as an order of magnitude for some
values of $m_\chi$. This feature corresponds to the coannihilation strip, which is shown
in Figure \ref{fig:mint} to dip into the $\stau$-LSP
excluded region near $m_{1/2} = 900$ GeV. The variation in the cross
section in this coannihilation strip region at low $m_{\chi}$ in panel (a) is due to
the separation of the coannihilation strip from the boundary of the $\stau$-LSP
region at low $m_{1/2}$. The cross
sections for points lying between the coannihilation strip and the
forbidden $\stau$-LSP region, where the relic density of neutralinos
is too low, are scaled down to reflect the fact that
in these cases the neutralinos can provide only a small fraction of the
cold dark matter in the Universe.

The second region lies within $80$ GeV $< m_{\chi} <
170$~GeV. In the case of the spin-dependent cross section, the cross
sections in this region are
clearly separated from those due to the coannihilation
strip. This second region of acceptable cross sections comes from the focus-point 
region which, for $M_{in} = M_{GUT}$, occurs at large $m_0$ and
small $m_{1/2}$. It should be noted that, if we were to consider values of
$m_0 > 2000$ GeV, the focus-point region would extend to larger
$(m_{1/2},m_0)$, so analogous focus-point cross sections would extend also
to larger $m_{\chi}$. In the focus-point region, the fact that the lightest neutralino
acquires substantial Higgsino components leads to an enhancement in the
spin-dependent cross section due to $Z$ exchange. Simultaneously, the scalar cross
section becomes dominated by neutral Higgs exchange as the neutralino
becomes Higgsino-like.

Panels (b), (c), and (d) show the neutralino-nucleon cross sections as
functions of the neutralino mass for $M_{in} = 10^{14}$, $10^{12}$, and
$10^{10}$ GeV, respectively. The changes in
the cross sections as $M_{in}$ is lowered may be understood by
referring to the corresponding $(m_{1/2},m_0)$ planes from
Figures \ref{fig:mint} and \ref{fig:mint2}.  When $M_{in} = 10^{14}$
GeV, the focus-point region becomes more prominent, separating from the
boundary of the region excluded by the electroweak vacuum condition.
In fact, for the portion of the $(m_{1/2},m_0)$ plane 
shown in Fig.~\ref{fig:mint}, the focus-point region extends to larger $m_{\chi}$ than the
coannihilation strip.  The two regions are seen as merged in panel (b-d). 

As we proceed to panel
(c), most of the $(m_{1/2},m_0)$ plane results in a relic density of
neutralinos that is within or below the cosmologically-preferred range.
As a result, there is a uniform distribution of possible cross sections up to
$m_{\chi} \approx 650$ GeV.  The
upper boundaries of the scalar and spin-dependent cross sections in
panel (c) come from regions in the plane where the relic density is
largest and $m_0$ is lowest, i.e., from the WMAP-preferred regions found
at low $m_0$. The continuous WMAP region that extends from the
$\stau$-LSP boundary to larger $m_0$ and $m_{1/2}$ is responsible for
this uniform upper limit for the cross sections for $m_{\chi} \lesssim
650$ GeV.  Near $m_{1/2} = 1100$ GeV, however, a new region of
preferred relic density emerges at lower $m_0$, leading to a bump in the
neutralino-nucleon cross sections that extends to the largest values
of $m_{\chi}$ considered here. 

This same behavior is observed in panel
(d), where $M_{in} = 10^{10}$ GeV.  The relic density of neutralinos
falls within the WMAP range only in a small region of the
$(m_{1/2},m_0)$ plane with $m_{1/2} >
1700$ GeV and is too small elsewhere, but similar increases
and decreases in the relic density where different annihilation
channels dominate are evident. We also point out that, since the relic density is lower than
the WMAP range over most of the plane for $M_{in} = 10^{12}$ and
$10^{10}$ GeV, we clearly see the maximum weak-scale value of $m_{\chi}$,
which corresponds to $m_{1/2} = 2000$ GeV, decrease
between panel (c) and panel (d). For $M_{in} < 10^{12}$ GeV, the LSP becomes
Higgsino-like, with $m_{\chi} \sim \mu$ and $\mu$ decreasing rapidly as
$M_{in}$ is lowered, as discussed in Section~\ref{sec:renorm}.  For
$\tbt = 10$, the cross sections excluded by CDMS come only from points in
the $(m_{1/2},m_0)$ plane that also fail the relaxed LEP Higgs constraint.

In Figure \ref{fig:cs50} we show the neutralino-nucleon cross sections
for $\tbt=50$ with $M_{in} = M_{GUT}$, $10^{15}$, $10^{14}$ and
$10^{12}$ GeV. Although the cosmologically-preferred regions of the
$(m_{1/2},$ $m_0)$ plane are somewhat different from those for
$\tbt=10$, the plots in Fig.~\ref{fig:cs50} look qualitatively
similar to those in Fig.~\ref{fig:cs10}. In panel (a) there
is a clear separation between the cross sections from the focus-point
region and those from the coannihilation strip and the beginning of the rapid-annihilation
funnel. Since the funnel region of acceptable relic density  pictured in panel (a) of 
Fig.~\ref{fig:mint50} extends to
$m_{1/2} \approx 1850$ GeV, we find values of the cross
sections out to $m_{\chi} \approx 850$ GeV. We note that some of the
scalar cross sections for $m_{\chi} \lesssim 200$ GeV that pass all
other constraints outlined above have been excluded by CDMS.

In panel (b), where $M_{in} = 10^{15}$ GeV, the two regions
are still distinct.  The lower
bulk of cross sections comes now from points inside the
fully-developed rapid-annihilation funnel, seen in panel (c) of Fig.~\ref{fig:mint50}. Again, we
note that had we extended our analysis to larger values of $m_{1/2}$
and $m_0$, acceptable cross sections would be found also at larger
$m_{\chi}$. 

When $M_{in}=10^{14}$ GeV, shown in panel (c), the upper funnel wall
has passed through the focus point, and only the lower funnel
wall remains.  Regions to the left of this wall in the $(m_{1/2},m_0)$
plane are essentially inside the funnel and have a very low relic
density of neutralinos, whereas the relic density is too large to the
right of the wall. Consequently, we see in panel (c) that, at low $m_{\chi}$, the
scalar and spin-dependent cross sections span several orders of magnitude.

As in the case when $\tbt = 10$, at low $M_{in}$ the relic
density of neutralinos falls within or below the WMAP range over all of
the $(m_{1/2},m_0)$ plane, so none
of the plane is excluded by the constraint on the relic density. This
is the case in Panel (d), where $M_{in} = 10^{12}$ GeV. The situation
remains unchanged as $M_{in}$ is further decreased.

We note that the scalar cross sections are generally larger at large $\tbt$. In fact, some of these cross sections are already
excluded by ZEPLIN-II  as well as CDMS II, which both probe WIMP-nucleon
scalar cross sections as low as a few $\times 10^{-7}$~pb \cite{zep2,cdms2}. A sensitivity of
$10^{-9}$~pb for $M_{\chi} \approx 100$ GeV is expected for
SuperCDMS Phase A with seven towers deployed \cite{supercdms}.  Many direct dark-matter
search experiments plan to use Xenon or Argon as an alternative target material for
which the sensitivity scales linearly with the detector mass. The
Argon Dark Matter exeriment (ArDM) expects to probe spin-independent
cross sections as low as $10^{-10}$~pb with a one-tonne detector
operating for one year \cite{ardm}. Results from direct detection experiments will
provide a useful complement to searches for SUSY signatures at colliders.

\section{Summary}

We have examined the impact of lowering the scale of unification of
the soft supersymmetry-breaking parameters of the CMSSM on
phenomenological, collider and cosmological constraints. In order to
carry out this study, we accounted for coannihilations involving the three lightest
neutralinos, the lighter chargino, and relevant sleptons and squarks.
We explored $\tbt
=10$ and $\tbt=50$, $A_0 \ne 0$, and a specific case similar to those
found in mirage-mediation models.  Intermediate unification scales
result in the appearance of a rapid-annihilation funnel even at
low $\tbt$, and the merging of this funnel and the focus-point
region as $M_{in}$ decreases. As the unification
scale is lowered below a critical value dependent on $\tbt$ and other factors, the relic density of neutralinos becomes too low to account
fully for the required relic density of cold
dark matter over all or nearly all of the $(m_{1/2},m_0)$ plane. These
values of $M_{in}$ are disfavored in the sense that there must be
another source of astrophysical cold dark matter in the universe.

We have also presented the neutralino-nucleon cross sections for
several values of $M_{in}$ at $\tbt =10$ and $\tbt=50$. We find that
the spin-independent neutralino-nucleon cross sections for regions of parameter space
favored by cosmology are beginning to be excluded by CDMS and other
direct detection WIMP searches, although viable cross sections span
several orders of magnitude. We look forward to stronger limits on the
spin-independent cross sections as direct-detection WIMP searches
become more sensitive in the near future.

The analysis in this paper has shown that lowering the scale of unification in
even the simplest CMSSM model may alter significantly the phenomenological expectations for 
both collider and non-collider experiments. It has also revealed novel effects in the
calculation of the relic neutralino density, such as the importance of multi-channel
neutralino and chargino coannihilation processes. However, we have done little more than
scratch the surface of possibilities since, for example, we have not considered in detail
scenarios with different values of $A_0$, let alone non-CMSSM scenarios or more realistic
mirage-mediation models. Another interesting and important question for
the future is the accuracy with which the effective unification scale could be estimated on
the basis of future collider experiments. We hope that this work will trigger future
studies of these and other related issues.

\section*{Acknowledgments}
\noindent 
The work of K.A.O. and P.S. was supported in part
by DOE grant DE--FG02--94ER--40823.

\appendix

\section{Neutralino and Chargino Coannihilations}

In most standard CMSSM scenarios, the LSP is
a bino-like neutralino in many of the
regions of parameter space relevant to cosmology, possibly with a significant
Higgsino admixture. When the relic density falls near the
range favoured by WMAP and other measurements, the neutralinos are
typically not degenerate, and therefore there is no opportunity for
coannihilations of the LSP with other neutralinos, or with
charginos, to bring the relic density down into the range preferred by
cosmology. The only case in which it is necessary to include
coannihilations involving neutralinos and charginos occurs when the
neutralino LSP is Higgsino-like, which arises when $\mu < M_1$, a situation that may arise
at large $m_0$ in the focus-point region of the GUT-scale CMSSM~\footnote{This situation may 
also occur in some
models with non-universal Higgs masses.}. In such a case it is possible
for the lightest and second-lightest neutralinos to be degenerate with
each other and with the lightest chargino. Thus, at large $m_0$ and small
$m_{1/2}$ in the GUT-scale CMSSM, coannihilations between the lightest and second lightest
neutralinos and with charginos must be included~\footnote{It should be noted
that $\stau$-$\chi$ coannihilations are known to be of general importance in the CMSSM, 
since they give rise to the so-called coannihilation
strip.}.

In GUT-less scenarios, the lightest neutralino becomes Higgsino-like at
low $M_{in}$, as discussed in Section \ref{sec:renorm}, so it is
necessary to include coannihilations involving the two lightest
neutralinos and the lightest chargino as discussed above.  
However, there is also a region of parameter space where additional coannihilations 
become significant. When the
LSP is mixed and nearly degenerate with the second lightest
neutralino and the lightest chargino, in some circumstances the third-lightest neutralino
may also be nearly degenerate. 

\begin{figure}
\begin{center}
\mbox{\epsfig{file=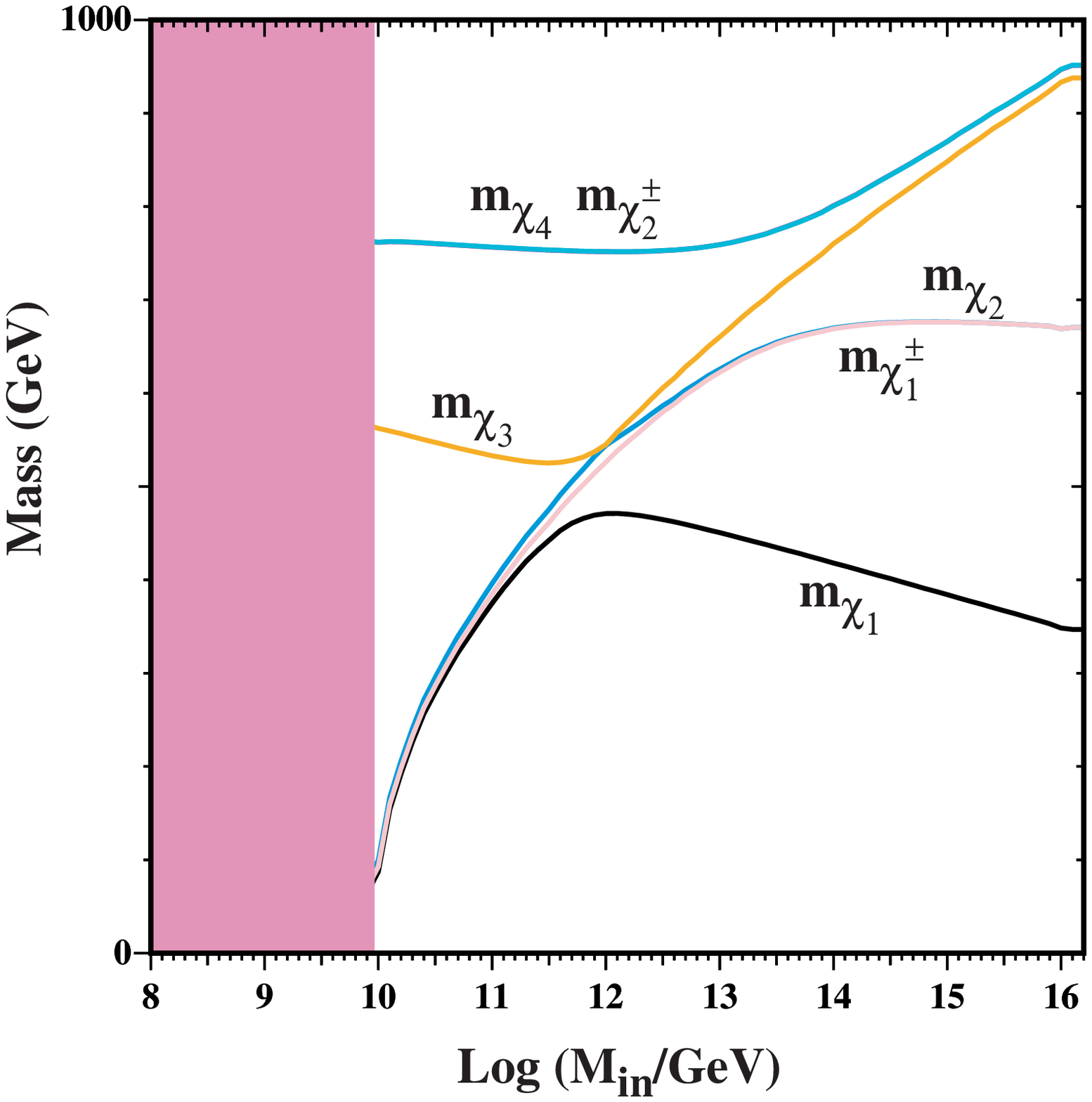,height=7cm}}
\mbox{\epsfig{file=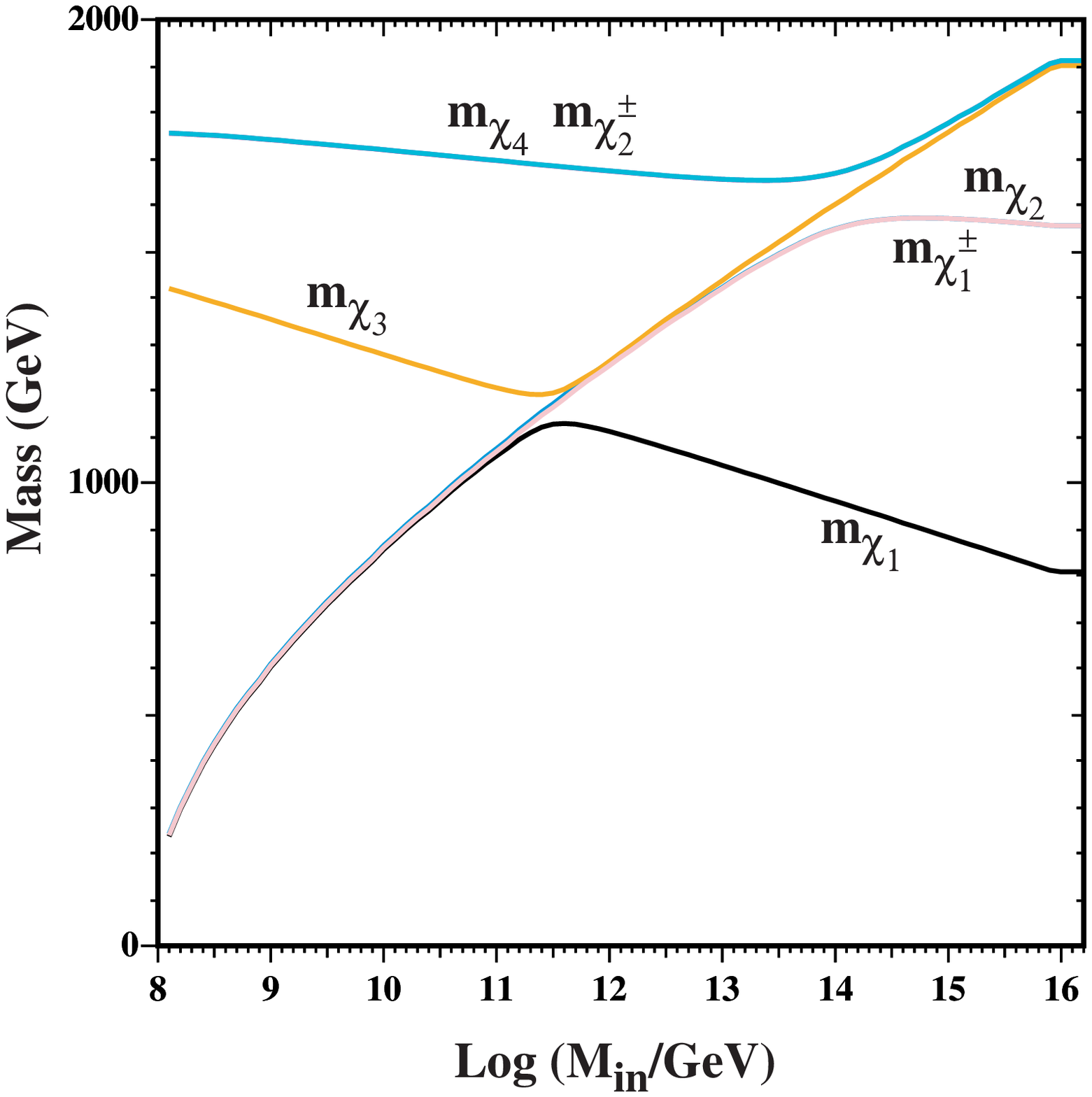,height=7cm}}
\end{center}
\begin{center}
\mbox{\epsfig{file=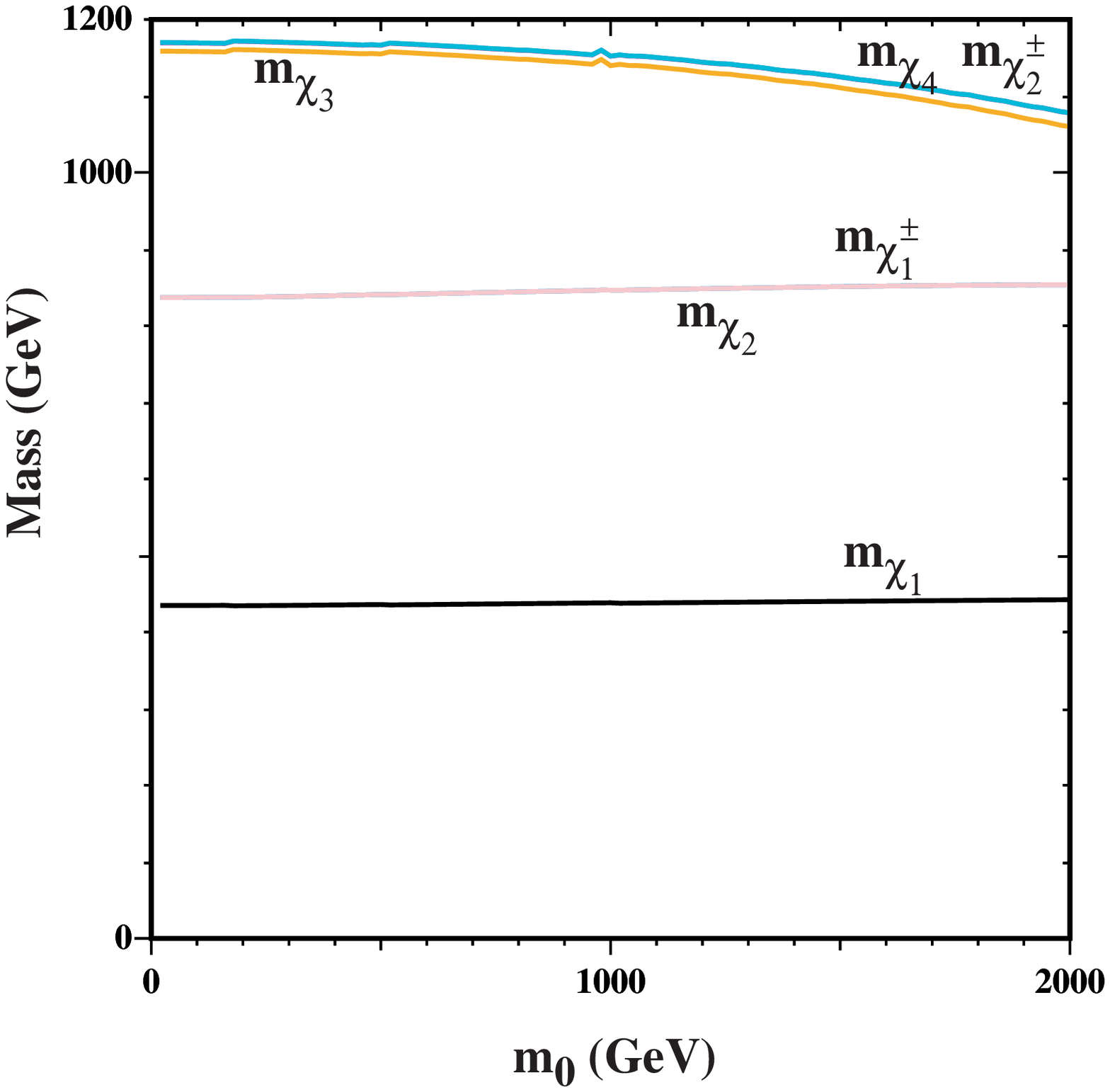,height=7cm}}
\mbox{\epsfig{file=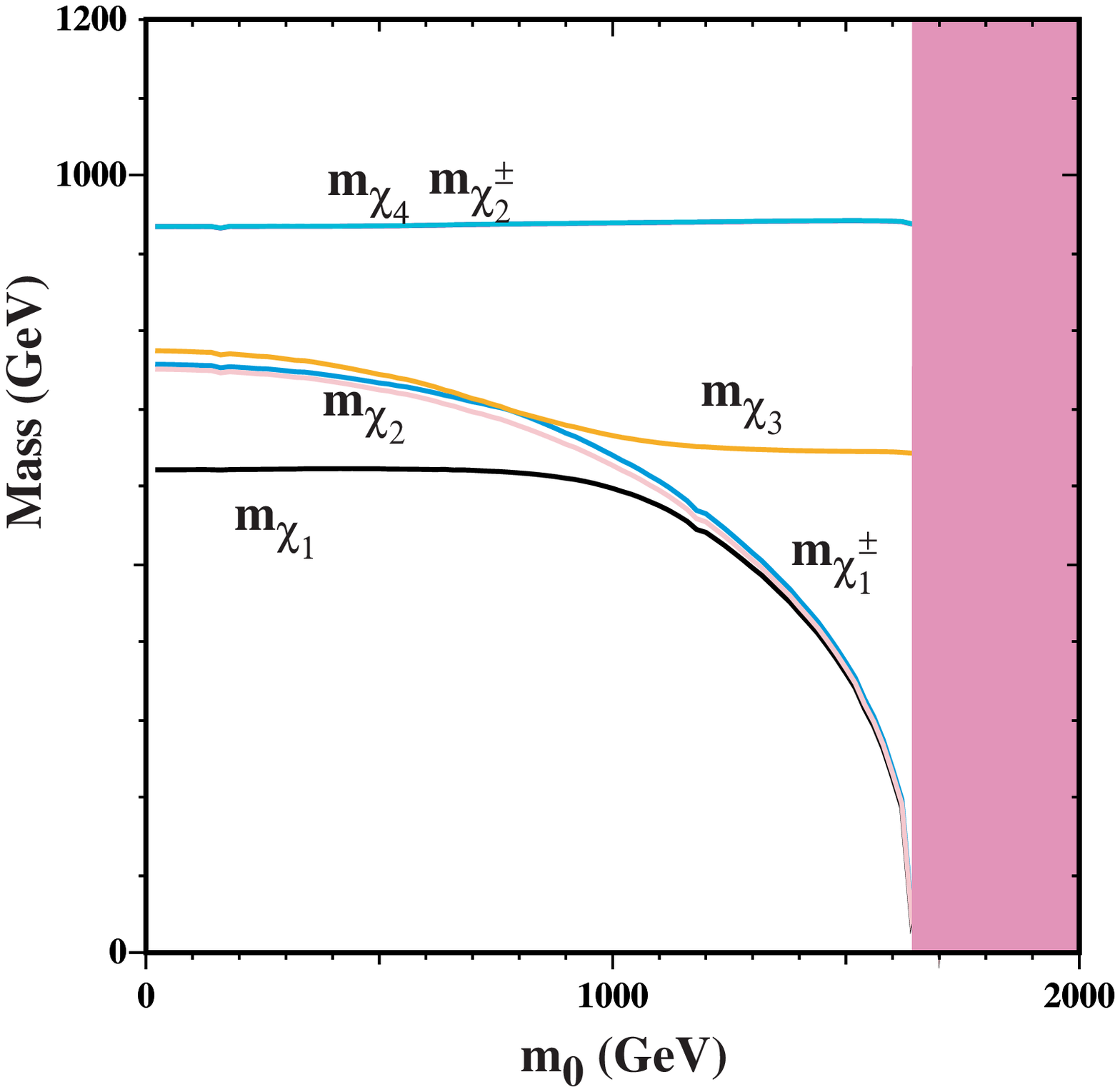,height=7cm}}
\end{center}
\caption{\it Panels (a) and (b) show the neutralino and chargino masses as functions of $M_{in}$
for the points $(m_{1/2},m_0) = (800,1000)$ and $(1800,1000)$ GeV,
respectively.  The bottom two panels show the neutralino and chargino masses as functions of
$m_0$ with $m_{1/2} = 1000$ GeV for (c) the GUT-scale CMSSM case and (d) the GUT-less case
with $M_{in} = 3 \times 10^{11}$ GeV.
}
\label{fig:neutralinos}
\end{figure}

In panels (a) and (b) of Figure \ref{fig:neutralinos}
we show the masses of all neutralinos and charginos as
functions of $M_{in}$ for two different points in the $(m_{1/2},m_0)$
plane. We recall from the discussion in Section \ref{sec:renorm} that the
slope of the curve describing the LSP mass is an indication of its
composition: when the neutralino mass increases as $M_{in}$ decreases,
it is gaugino-like, and when it decreases as $M_{in}$ decreases,
tracking $|\mu|$, it is Higgsino-like. Panels (a) and (b) show that the masses of the LSP, the second
lightest neutralino and the chargino are nearly degenerate when the LSP is
Higgsino-like, indicating the necessity of including coannihilations
involving all three states. Moreover, just at the point where the LSP changes
from bino-like to Higgsino-like, the mass of the third-lightest
neutralino dips down near the masses of the two lighter
neutralinos and the chargino. Panels (c) and (d) compare the masses of the
neutralinos and charginos as functions of $m_0$ for the CMSSM case
with GUT-scale universality and $M_{in} = 3 \times 10^{11}$ GeV for
fixed $m_{1/2} = 1000$ GeV. One can see in panel (c) that, for the GUT-scale case, there
is no degeneracy of the LSP with other neutralinos or charginos. The
LSP is strongly bino-like, and therefore its mass is related to
$m_{1/2}$, as in (\ref{gaugino}), with only a very weak
dependence on $m_0$ through higher-order corrections. The same
scenario is shown in panel (d) for $M_{in} = 3 \times 10^{11}$ GeV. In this case, however, 
we see that mass degeneracies are apparent over a wide range of values of $m_0$.


For some values of
$M_{in}$, the near-degeneracy of the third-lightest neutralino with
lighter neutralinos and the chargino occurs precisely where the relic
density of neutralinos is near the cosmologically preferred value.
For example, for $M_{in} = 10^{12}$ GeV and $\tbt = 10$, the shape and location of the WMAP strip
running through $(1500,1000)$ GeV can shift by as much as 200 GeV in $m_0$
if coannihilations are not properly included. 

Thus, the calculations of the coannihilation processes that previously were
included for the second-lightest neutralino have here been calculated
also for the third-lightest neutralino, including those of $\chi_2$ with
$\chi_3$.  Table \ref{coan} shows the initial states for all the
calculated annihilations and coannihilations of neutralinos and
charginos used in the analysis here.  In addition to those outlined
below, coannihilations of all these neutralino and chargino species with sfermions were calculated,
as well as the corresponding sfermion-antisfermion annihilation processes.

\begin{table}[htb]
\begin{center}
\begin{tabular}{|c|c|c|}
\hline
$\chi_1 \chi_1$ & $\chi_1 \chi_2$ & $\chi_1 \chi_+$ \\
\hline
$\chi_2 \chi_2$ & $\chi_2 \chi_3$ & $\chi_2 \chi_+$ \\
\hline
$\chi_3 \chi_3$ & $\chi_3 \chi_1$ & $\chi_3 \chi_+$ \\
\hline
 & $\chi_+ \chi_-$ &    \\
\hline
\end{tabular}
\caption{\it Initial states of interactions included here in the calculation of the
relic cold dark matter density, where $\chi_1$ is the LSP and
$\chi_{2(3)}$ is the second (third) -lightest neutralino.
\label{coan}}
\end{center}
\end{table}

As noted in the main text, we take the opportunity in this paper to improve on our
previous treatment of the rapid-annihilation region and to correct certain coding
inaccuracies which, however, have no visible effects on the results we present.

\end{document}